\documentclass[10pt,twocolumn]{IEEEtran}
\usepackage{amssymb,amsmath,mathrsfs}
\usepackage{color,array,times}
\usepackage{epsfig,epstopdf,textcomp}
\begin{document}

\title{Design of Robust Adaptive Beamforming Algorithms Based on Low-Rank and Cross-Correlation Techniques}

\author{Hang Ruan and Rodrigo C. de Lamare}
\maketitle

\begin{abstract}
This work presents cost-effective low-rank techniques for designing robust adaptive beamforming (RAB) algorithms. The proposed algorithms are based on the exploitation of the cross-correlation between the array observation data and the output of the beamformer. Firstly, we construct a general linear equation considered in large dimensions whose solution yields the steering vector mismatch. Then, we employ the idea of the full orthogonalization method (FOM), an orthogonal Krylov subspace based method, to iteratively estimate the steering vector mismatch in a reduced-dimensional subspace, resulting in the proposed orthogonal Krylov subspace projection mismatch estimation (OKSPME) method. We also devise adaptive algorithms based on stochastic gradient (SG) and conjugate gradient (CG) techniques to update the beamforming weights with low complexity and avoid any costly matrix inversion. The main advantages of the proposed low-rank and mismatch estimation techniques are their cost-effectiveness when dealing with high dimension subspaces or large sensor arrays. Simulations results show excellent performance in terms of the output signal-to-interference-plus-noise ratio (SINR) of the beamformer among all the compared RAB methods.
\end{abstract}


\section{introduction}

Adaptive beamforming has been one of the most important research areas in sensor array signal processing. It has also been recognized that traditional adaptive beamformers are extremely sensitive to environmental uncertainties or steering vector mismatches, which may be caused by many different factors (e.g., imprecise antenna size calibration, signal pointing errors or local scattering). Furthermore, some radar systems in advanced applications require antenna arrays with a very large number of sensor elements in highly dynamic environments, which leads to the increase of computational complexity and the decrease of the convergence rate for computing the parameters of the beamformer.

\subsection{{Prior and Related Work}}

In order to mitigate the effects of uncertainties on adaptive
beamformers, robust adaptive beamforming (RAB) techniques have been
developed. Popular approaches include worst-case optimization
\cite{r2,r9}, diagonal loading \cite{r4}, and eigen-subspace
decomposition and projection techniques \cite{r6,r8,r11}. However,
these RAB approaches have some limitations such as their ad hoc
nature, high probability of subspace swap at low signal-to-noise
ratio (SNR) \cite{r3} and high computational cost due to online
optimization or subspace decomposition techniques.

Furthermore, in the case of large sensor arrays the above mentioned
RAB methods may encounter problems for their application. This is
because in these RAB algorithms, a cubic or greater computational
cost is required to compute the beamforming parameters. Therefore,
dimensionality reduction (or rank-reduction) methods
(\cite{r15}-\cite{r31}, \cite{r33}-\cite{r36}) have been employed
and developed to reduce the complexity and improve the convergence
rate.

In the recent years, great efforts have been devoted to the
investigation of dimensionality reduction techniques
\cite{scharf,bar-ness,pados99,reed98,hua,goldstein,santos,qian,delamarespl07,xutsa,delamaretsp,kwak,xu&liu,delamareccm,wcccm,delamareelb,jidf,delamarecl,delamaresp,delamaretvt,jioel,delamarespl07,delamare_ccmmswf,jidf_echo,delamaretvt10,delamaretvt2011ST,delamare10,fa10,lei09,ccmavf,lei10,jio_ccm,ccmavf,stap_jio,zhaocheng,zhaocheng2,arh_eusipco,arh_taes,dfjio,rdrab,dcg_conf,dcg,dce,drr_conf,dta_conf1,dta_conf2,dta_ls,song,wljio,barc,jiomber,saalt,alrdoa}.
 for RAB. The
beamspace approach of \cite{r23} projects the data onto a lower
dimension subspace by using a beamspace matrix, whose columns are
determined by linearly independent constrained optimization
problems. A more effective approach (i.e.,
\cite{r16}-\cite{r20},\cite{r24},\cite{r25}) is based on
preprocessing the array observation data using a Krylov subspace.
However, there are different ways to generate the Krylov subspace
and the choice usually depends on the cost and the performance. The
Arnoldi method \cite{r12,r13,r21} and the Lanczos iterations
\cite{r12,r13,r15} are typical approaches used to generate
orthogonal Krylov subspaces, whereas \cite{r19} also introduces a
method to generate non-orthogonal ones. However, the main challenge
in these techniques is the model order determination. Specifically,
the model order must be properly chosen to ensure robustness to
over-determination of the system model order \cite{r17}. Another
effective approach to dimensionality reduction is the joint
iterative optimization (JIO) \cite{r26}-\cite{r31} techniques,
\cite{r33}-\cite{r36}, which employ a subspace projection matrix and
jointly and iteratively optimize the bases of the subspace and the
beamformer weights. The work in \cite{r27} has developed a recursive
least squares (RLS) adaptive algorithm based on widely-linear
processing using the JIO technique. The study in \cite{r29} has
devised efficient stochastic gradient (SG) and RLS RAB algorithms
from a modified JIO (MJIO) scheme.

\subsection{{Contributions}}

In this work, we propose and study novel RAB algorithms that are
based on low-rank and cross-correlation techniques. In the proposed
techniques, we exploit the prior knowledge that the steering vector
mismatch of the desired signal is located within an angular sector
which is assumed known. The proposed algorithms are based on the
exploitation of the cross-correlation between the array observation
data and the output of the beamformer, which avoids costly
optimization procedures. We firstly construct a linear system
(considered in high dimension) involving the mismatched steering
vector and the statistics of the sampled data. Then we employ an
iterative full orthogonalization method (FOM) \cite{r12,r13} to
compute an orthogonal Krylov subspace whose model order is
determined by both the minimum sufficient rank \cite{r17}, which
ensures no information loss when capturing the signal of interest
(SoI) with interferers, and the execute-and-stop criterion of FOM
\cite{r12,r13}, which automatically avoids overestimating the number
of bases of the computed subspace. The estimated vector that
contains the cross-correlation between the array observation data
and the beamformer output is projected onto the Krylov subspace, in
order to update the steering vector mismatch, resulting in the
proposed orthogonal Krylov subspace projection mismatch estimation
(OKSPME) method.

Furthermore, based on the OKSPME method, we have also devised
adaptive stochastic gradient (SG), conventional conjugate gradient
(CCG) and modified conjugate gradient (MCG) algorithms derived from
the proposed optimization problems to reduce the cost for computing
the beamforming weights for large sensor arrays
\cite{mmimo,wence,Costa,delamare_ieeproc,TDS_clarke,TDS_2,armo,buffer,switch_int,switch_mc,smce,TongW,jpais_iet,TARMO,keke1,kekecl,keke2,Tomlinson,dopeg_cl,peg_bf_iswcs,gqcpeg,peg_bf_cl,Harashima,mbthpc,zuthp,rmbthp,Hochwald,BDVP},\cite{delamare_mber,rontogiannis,delamare_itic,stspadf,choi,stbcccm,FL11,jio_mimo,peng_twc,spa,spa2,jio_mimo,P.Li,jingjing,did,bfidd,mbdf}.
, resulting in the proposed OKSPME-SG, OKSPME-CCG and OKSPME-MCG RAB
algorithms. We remark that the steering vector is also estimated and
updated using the CG-based recursions to produce an even more
precise estimate. Derivations of the proposed algorithms are
presented and discussed along with an analysis of their
computational complexity.

Moreover, we develop an analysis of the mean squared error (MSE) between the estimated and the actual steering vectors for the general approach of using a presumed angular sector associated with subspace projections. This analysis mathematically describes how precise the steering vector mismatch can be estimated. Upper and lower bounds are derived and compared with the approach in \cite{r6}. Another analysis on the computational complexity of the proposed and existing algorithms is also provided.

In the simulations, we consider local scattering scenarios (both coherent and incoherent) to model the mismatch effects. We also study the performance of the proposed algorithms by testing the output signal-to-interference-plus-noise ratio (SINR) of the beamformer with respect to training snapshots and different input SNRs. The number of sensor elements and interferers is also varied and compared in each scenario to provide a comprehensive performance study.
In summary, the contributions of this work are:
\begin{itemize}
\item The proposed OKSPME RAB method.
\item The development of the modified SG and CG type OKSPME RAB algorithms.
\item An analysis of the computational complexity and the MSE performance of the proposed and existing RAB algorithms.
\end{itemize}

The remaining sections of this paper are organized as follows: The system model and problem statement are described in Section II. Section III introduces the proposed OKSPME method, whereas Section IV introduces the proposed robust adaptive algorithms. Section V provides the MSE analysis of the steering vector estimation and the complexity analysis. Section VI presents and discusses the simulation results. Section VII gives the conclusion.

\section{System Model and Problem Statement}

Let us consider a linear antenna array of $M$ sensors and $K$ narrowband signals which impinge on the array. The data received at the $i$th snapshot can be modeled as
\begin{equation}
{\bf x}(i)={\bf A}({\boldsymbol\theta}){\bf s}(i)+{\bf n}(i), \label{eq1}
\end{equation}
where ${\bf s}(i) \in {\mathbb C}^{K \times 1}$ are uncorrelated source signals,
${\boldsymbol\theta}=[{\theta}_1,\dotsb,{\theta}_K]^T \in {\mathbb
R}^K$ is a vector containing the directions of arrival (DoAs) and $[.]^T$ denotes the transpose, ${\bf
A}({\boldsymbol \theta})=[{\bf a}({\theta}_1)+{\bf e}, \dotsb, {\bf
a}({\theta}_K)]=[{\bf a}_1, \dotsb, {\bf a}_K] \in {\mathbb C}^{M \times K}$ is the matrix which
contains the steering vector for each DoA and ${\bf e}$ is the
steering vector mismatch of the desired signal, ${\bf n}(i) \in
{\mathbb C}^{M \times 1}$ is assumed to be complex circular Gaussian noise
with zero mean and variance ${\sigma}^2_n$. The beamformer output is given by
\begin{equation}
y(i)={\bf w}^H{\bf x}(i), \label{eq2}
\end{equation}
where ${\bf w}=[w_1,\dotsb,w_M]^T \in {\mathbb C}^{M\times1}$ is the
beamformer weight vector, where $({\cdot})^H$ denotes the Hermitian
transpose. The optimum beamformer is computed by maximizing the
SINR and is given by
\begin{equation}
SINR=\frac{{\sigma}^2_1{\lvert{\bf w}^H{\bf a}_1\rvert}^2}{{\bf w}^H{\bf R}_{\textcolor{red}{I+N}}{\bf w}}. \label{eq3}
\end{equation}
where ${\sigma}^2_1$ is the desired signal power, ${\bf R}_{\textcolor{red}{I+N}}$ is the interference-plus-noise covariance (INC) matrix. The problem of maximizing the SINR in \eqref{eq3} can be cast as the following optimization problem:
\begin{equation}
\begin{aligned}
& \underset{\bf w} {\text{minimize}}
&& {\bf w}^H{\bf R}_{\textcolor{red}{I+N}}{\bf w} \\
& \text{subject to} && {\bf w}^H{\bf a}_1=1, \label{eq4}
\end{aligned}
\end{equation}
which is known as the MVDR beamformer or Capon beamformer \cite{r1,r4}. The optimum weight vector is given by $${\bf w}_{opt}=\frac{{{\bf R}^{-1}_{\textcolor{red}{I+N}}{\bf a}_1}}{{\bf a}_1^H{{\bf R}^{-1}_{\textcolor{red}{I+N}}{\bf a}_1}}$$. Since ${\bf R}_{\textcolor{red}{I+N}}$ is usually unknown in practice, it can be estimated by the sample covariance matrix (SCM) of the received data as
\begin{equation}
\hat{\bf R}(i)=\frac{1}{i}\sum\limits_{k=1}^i{\bf x}(k){{\bf x}^H}(k). \label{eq5}
\end{equation}

Using the SCM for directly computing the weights will lead to the sample matrix inversion (SMI) beamformer ${\bf w}_{SMI}=\frac{\hat{\bf R}^{-1}{\bf a}_1}{{\bf a}_1^H\hat{\bf R}^{-1}{\bf a}_1}$. However, the SMI beamformer requires a large number of snapshots to converge and is sensitive to steering vector mismatches \cite{r2,r3}. As previously mentioned, most of the conventional and existing RAB algorithms are computationally costly especially when encountering arrays with a very large number of sensors. Therefore, the RAB design problem we are interested in solving includes the following aspects:
\begin{itemize}
\item To design cost-efficient algorithms that are robust against uncertainties and values of SNRs and interferers in the presence of uncertainties in the steering vector of a desired signal.
\item The proposed algorithms must preserve their robustness and low-complexity features for large sensor arrays.
\end{itemize}

\section{Proposed OKSPME Method}

In this section, the proposed OKSPME method is introduced. This method aims to construct a linear system involving only known or estimated statistics and then projects an estimated cross-correlation vector between the array observation data and the beamformer output onto an orthogonal Krylov subspace, in order to update the steering vector mismatch with reduced complexity. The SCM of the array observation data is estimated by \eqref{eq5}. The cross-correlation vector between the array observation data and the beamformer output can be expressed as ${\bf d}=E[{\bf x}y^*]$ (where $[.]^*$ denotes complex conjugation) or equivalently as
\begin{equation}
{\bf d}=E[({\bf A}{\bf s}+{\bf n})({\bf A}{\bf s}+{\bf n})^H{\bf w}]. \label{eq6}
\end{equation}
\textcolor{red}{
Assuming that ${\lvert{{\bf a}_k^H{\bf w}}\rvert}\ll{\lvert{{\bf a}_1^H{\bf w}}\rvert}$ for $k=2,\dotsb,K$ and all signals have zero mean, the cross-correlation vector ${\bf d}$ can be rewritten as
\begin{equation}
{\bf d}=E[({\bf A}{\bf s}+{\bf n})(s_1^*{\bf a}_1^H{\bf w}+{\bf n}^H{\bf w})]. \label{eq7}
\end{equation}
Note that we also assume that the desired signal is statistically independent from the interferers and the noise, i.e., $E[s_k{s_1^*}]=0$ and $E[{s_k{\bf a}_k}s_1^*{\bf a}_1^H{\bf w}]=0$ for $k=2,\dotsb,K$. With this assumption the desired signal power is not statistically affected by the interference and \eqref{eq7} can be rewritten as
}
\begin{equation}
{\bf d}=E[{{\sigma}_1}^2{\bf a}_1^H{\bf w}{\bf a}_1+{\bf n}{\bf n}^H{\bf w}], \label{eq8}
\end{equation}
which can be estimated by the sample cross-correlation vector (SCV) given by
\begin{equation}
\hat{\bf d}(i)=\frac{1}{i}\sum\limits_{k=1}^i{\bf x}(k){y^*}(k). \label{eq9}
\end{equation}

\subsection{{Desired Signal Power Estimation}}

In this subsection, we describe an iterative method for the desired signal power (${\sigma}^2_1$) estimation based on our prior work in \cite{r32}, which can be accomplished by directly using the desired signal steering vector. \textcolor{red}{Alternatively, a designer can employ a maximum likelihood (ML) or a minimum variance (MV) estimator for computing the desired signal power. However, the approach described here has lower complexity than the ML and the MV estimators.} In the adopted method, we need to choose an initial guess for the steering vector mismatch within the presumed angular sector, say $\hat{\bf a}_1(0)$ and set $\hat{\bf a}_1(1)=\hat{\bf a}_1(0)$. By adding the snapshot index $i$, we can rewrite the array observation data as
\begin{equation}
{\bf x}(i)=\hat{\bf a}_1(i)s_1(i)+\sum\limits_{k=2}^K{\bf a}_k(i)s_k(i)+{\bf n}(i), \label{eq10}
\end{equation}
\textcolor{red}{where $\hat{\bf a}_1(0)$ and $\hat{\bf a}_1(i)$ ($i=1,2,\dotsb$) designate the initial guess of the steering vector and its estimate at the $i$th snapshot, respectively.}

Pre-multiplying the above equation by $\hat{\bf a}_1^H(i)$ we have
\begin{equation}
\hat{\bf a}_1^H(i){\bf x}(i)=\hat{\bf a}_1^H(i)\hat{\bf a}_1(i)s_1(i)+\sum\limits_{k=2}^K\hat{\bf a}_1^H(i){\bf a}_k(i)s_k(i)+{\bf n}(i). \label{eq-add}
\end{equation}
Here we assume that each of the interferers is orthogonal or approximately orthogonal to the desired signal. Specifically, the steering vector of each of the interferers is orthogonal ($\hat{\bf a}_1^H(i){\bf a}_k(i)=0$, $k=2,3,\dotsb,K$), or approximately orthogonal ($\hat{\bf a}_1^H(i){\bf a}_k(i)\ll\hat{\bf a}_1^H(i)\hat{\bf a}_1(i)$, $k=2,3,\dotsb,K$) to the desired signal steering vector (i.e., $\hat{\bf a}_1(i)$), so that $\hat{\bf a}_1^H(i){\bf a}_k(i)$ ($k=2,3,\dotsb,K$) approaches zero and the term $\sum\limits_{k=2}^K\hat{\bf a}_1^H(i){\bf a}_k(i)s_k(i)$ in \eqref{eq-add} can be neglected, resulting in
\begin{equation}
\hat{\bf a}_1^H(i){\bf x}(i)=\hat{\bf a}_1^H(i)\hat{\bf a}_1(i)s_1(i)+\hat{\bf a}_1^H(i){\bf n}(i). \label{eq11}
\end{equation}

Taking the expectation of $|\hat{\bf a}_1^H(i){\bf x}(i)|^2$, we obtain
\begin{multline}
E[|\hat{\bf a}_1^H(i){\bf x}(i)|^2]=E[(\hat{\bf a}_1^H(i)\hat{\bf a}_1(i)s_1(i)+\hat{\bf a}_1^H(i){\bf n}(i))^* \\ (\hat{\bf a}_1^H(i)\hat{\bf a}_1(i)s_1(i)+\hat{\bf a}_1^H(i){\bf n}(i))]. \label{eq12}
\end{multline}

Assuming that the noise is statistically independent from the desired signal, then we have
\begin{multline}
E[|\hat{\bf a}_1^H(i){\bf x}(i)|^2]=|\hat{\bf a}_1^H(i)\hat{\bf a}_1(i)|^2E[|s_1(i)|^2] \\
+\hat{\bf a}_1^H(i)E[{\bf n}(i){\bf n}^H(i)]\hat{\bf a}_1(i), \label{eq13}
\end{multline}
where $E[{\bf n}(i){\bf n}^H(i)]$ represents the noise covariance
matrix ${\bf R}_n(i)$ that can be replaced by ${\sigma}^2_n{\bf I}_M$,
where the noise variance ${\sigma}^2_n$ can be easily estimated by a specific estimation
method. A possible approach is to use a Maximum Likelihood (ML) based method as in \cite{r14}. Replacing the desired signal power $E[|s_1(i)|^2]$ and the noise variance ${\sigma}^2_n$ by their estimates
$\hat{\sigma}^2_1(i)$ and $\hat{\sigma}^2_n(i)$, respectively, we obtain
\begin{equation}
\hat{\sigma}^2_1(i)=\frac{|{{\hat{\bf a}_1}^H}(i){\bf x}(i)|^2-|{{\hat{\bf a}_1}^H}(i){\hat{\bf a}_1}(i)|\hat{\sigma}^2_n(i)}{|{{\hat{\bf a}_1}^H}(i){\hat{\bf a}_1}(i)|^2}. \label{eq14}
\end{equation}

The expression in \eqref{eq14} has a low complexity (${\mathcal O}(M)$) and can be directly implemented if the desired signal steering vector and the noise level are accurately estimated.

\subsection{{Orthogonal Krylov Subspace Approach for Steering Vector Mismatch Estimation}}

An orthogonal Krylov subspace strategy is proposed in order to estimate the mismatch with reduced cost and deal with situations in which the model order is time-varying. Our idea is based on constructing a linear system, which considers the steering vector mismatch as the solution, and solving it by using an iterative Krylov subspace projection method. To this end, consider a general high-dimensional linear system model given by
\begin{equation}
{\bf B}{\bf a}_1={\bf b}, \label{eq15}
\end{equation}
where ${\bf B} \in {\mathbb C}^{M \times M}$ and ${\bf b} \in {\mathbb C}^{M \times 1}$. Then we need to express ${\bf B}$ and ${\bf b}$ only using available information (known statistics or estimated parameters), so that we can solve the linear system with the Krylov subspace of order $m$ ($m \ll M$) described by
\begin{equation}
{\bf K}_m={\textit span}\{{\bf b}, {\bf B}{\bf b}, {\bf B}^2{\bf b}, \dotsb, {\bf B}^m{\bf b}\}. \label{eq16}
\end{equation}

Taking the complex conjugate of \eqref{eq11}, we have
\begin{equation}
{\bf x}^H(i)\hat{\bf a}_1(i)=\hat{\bf a}_1^H(i)\hat{\bf a}_1(i)s_1^*(i)+{\bf n}^H(i)\hat{\bf a}_1(i). \label{eq17}
\end{equation}
Pre-multiplying both sides of \eqref{eq17} by the terms of \eqref{eq10}, \textcolor{red}{then adding an extra term $\delta{\bf I}\hat{\bf a}_1(i)$ (where $\delta$ is a small positive number defined by the user) and
simplifying the terms, we obtain
\begin{multline}
({\bf x}(i){\bf x}^H(i)+\delta{\bf I})\hat{\bf a}_1(i)=\hat{\bf a}_1(i)\hat{\bf a}_1^H(i)\hat{\bf a}_1(i)s_1(i)s_1^*(i) \\
+{\bf n}(i){\bf n}^H(i)\hat{\bf a}_1(i)+\delta\hat{\bf a}_1(i). \label{eq18}
\end{multline}
Replacing ${\bf x}(i){\bf x}^H(i)+\delta{\bf I}$ by $\hat{\bf R}(i)$, $s_1(i)s_1^*(i)$ by $\hat{\sigma}_1^2(i)$ and ${\bf n}(i){\bf n}^H(i)$ by $\hat{\sigma}_n^2(i){\bf I}_M$, we obtain
\begin{equation}
\hat{\bf R}(i)\hat{\bf a}_1(i)=\underbrace{\hat{\bf a}_1(i)\hat{\bf a}_1^H(i)\hat{\bf a}_1(i)\hat{\sigma}_1^2(i)+(\hat{\sigma}_n^2(i)+\delta)\hat{\bf a}_1(i)}_{\hat{\bf b}(i)}, \label{eq19}
\end{equation}
in which by further defining the expression on the right-hand side as $\hat{\bf b}(i)$, we can rewrite \eqref{eq19} as
\begin{equation}
\hat{\bf R}(i)\hat{\bf a}_1(i)=\hat{\bf b}(i). \label{eq20}
\end{equation}
}

As can be seen \eqref{eq20} shares the same form as the linear
system of equations in \eqref{eq15} and $\hat{\bf b}(i)$ can be
expressed in terms of $\hat{\bf a}_1(i)$,
$\hat{\sigma}_1^2(i)$ and $\hat{\sigma}_n^2(i)$ whereas $\hat{\bf R}(i)$ can be estimated by \eqref{eq5}. In the following step, we employ the Arnoldi-modified Gram-Schmidt algorithm from the FOM method \cite{r12,r13} associated with the minimum sufficient rank criterion discussed in \cite{r17} to compute an orthogonal Krylov subspace. We define a residue vector to represent the estimation error in the $i$th snapshot as
\begin{equation}
\hat{\bf r}(i)=\hat{\bf b}(i)-\hat{\bf R}(i)\hat{\bf a}_1(i), \label{eq21}
\end{equation}
and let
\begin{equation}
{\bf t}_1(i)=\frac{\hat{\bf r}(i)}{\lVert{\hat{\bf r}(i)}\rVert}. \label{eq22}
\end{equation}

Then the Krylov subspace bases can be computed using the modified Arnoldi-modified Gram-Schmidt algorithm as in Table \ref{table1} (the snapshot index $i$ is omitted here for simplicity).

\begin{table}
\begin{center}
\caption{Arnoldi-modified Gram-Schmidt algorithm}
\begin{tabular}{l}
\hline
For $j=1,2,\dotsb$ do: \\
~~~~Compute ${\bf u}_j=\hat{\bf R}{\bf t}_j$ \\
~~~~For $l=1,2,\dotsb,j$, do: \\
~~~~~~~~$h_{l,j}=<{\bf u}_j,{\bf t}_l>$ \\
~~~~~~~~${\bf u}_j={\bf u}_j-h_{l,j}{\bf t}_l$ \\
~~~~End do. \\
~~~~Compute $h_{j,j+1}=\lVert{{\bf u}_j}\rVert$. \\
~~~~If $h_{j,j+1}=0$ or $j\geq K+1$, \\
~~~~~~~~set $m=j$; \\
~~~~~~~~break; \\
~~~~Else compute ${\bf t}_{j+1}=\frac{{\bf u}_j}{h_{j,j+1}}$. \\
End do. \\
\hline
\end{tabular} \label{table1}
\end{center}
\end{table}

In Table \ref{table1}, $<,>$ denotes the inner product and the parameters $h_{l,j}$ ($l,j=1,2,\dotsb,m$) are real-valued coefficients, the model order is determined once if one of the following situations is satisfied:
\begin{itemize}
\item The execute-and-stop criterion of the original Arnoldi-modified Gram-Schmidt algorithm is satisfied (i.e., $h_{j,j+1}=0$).
\item The minimum sufficient rank for addressing the SoI and interferers is achieved (i.e., $j\geq K+1$, where $K$ is the number of signal sources), so that no more subspace components are necessary for capturing the SoI from all the existing signal sources.
\end{itemize}

Now by inserting the snapshot index, we have
\begin{equation}
\hat{\bf T}(i)=[{\bf t}_{1}(i), {\bf t}_{2}(i), \dotsb, {\bf t}_{m}(i)], \label{eq23}
\end{equation}
and the Krylov subspace projection matrix is computed by
\begin{equation}
\hat{\bf P}(i)=\hat{\bf T}(i)\hat{\bf T}^H(i). \label{eq24}
\end{equation}

It should be emphasized that the Krylov subspace matrix $\hat{\bf T}(i)$ obtained here is constructed by starting with the residue vector $\hat{\bf r}(i)$. In other words, $\hat{\bf T}(i)$ is constructed with the estimation error of the steering vector. In order to extract the estimation error information and use it to update the steering vector mismatch, we can project the SCV $\hat{\bf d}(i)$ in \eqref{eq9} onto $\hat{\bf P}(i)$ and add the estimation error to the current estimate of $\hat{\bf a}_1(i)$ as
\begin{equation}
\hat{\bf a}_1(i+1)=\hat{\bf a}_1(i)+\frac{\hat{\bf P}(i)\hat{\bf d}(i)}{\lVert\hat{\bf P}(i)\hat{\bf d}(i)\rVert}. \label{eq25}
\end{equation}

\subsection{{INC Matrix and Beamformer Weight Vector Computation}}

Since we have estimated both the desired signal power $\hat{\sigma}^2_1(i)$ and the mismatched steering vector in the previous subsections, the INC matrix can be obtained by subtracting the desired signal covariance matrix out from the SCM as
\begin{equation}
\hat{\bf R}_{\textcolor{red}{I+N}}(i)=\hat{\bf R}(i)-\hat{\sigma}^2_1(i)\hat{\bf a}_1(i)\hat{\bf a}^H_1(i). \label{eq26}
\end{equation}

The beamformer weight vector is computed by
\begin{equation}
\hat{\bf w}(i)=\frac{{\hat{\bf R}^{-1}_{\textcolor{red}{I+N}}}(i)\hat{\bf a}_1(i)}{\hat{\bf a}^H_1(i)\hat{\bf R}^{-1}_{\textcolor{red}{I+N}}(i)\hat{\bf a}_1(i)}, \label{eq27}
\end{equation}
which has a computationally costly matrix inversion $\hat{\bf R}^{-1}_{\textcolor{red}{I+N}}(i)$. The proposed OKSPME method is summarized in Table \ref{table2}. In the next section, we will introduce adaptive algorithms to avoid matrix inversions and reduce the complexity.

\begin{table}
\small
\begin{center}
\caption{Proposed OKSPME method}
\begin{tabular}{l}
\hline
Initialization: \\
$\hat{\bf w}(1)={\bf 1}$; \\
Choose an initial guess $\hat{\bf a}_1(0)$ within the sector and set $\hat{\bf a}_1(1)=\hat{\bf a}_1(0)$; \\
For each snapshot $i=1,2,\dotsb$: \\
~~~~$\hat{\bf R}(i)=\frac{1}{i}\sum\limits_{k=1}^i{\bf x}(k){{\bf x}^H}(k)$ \\
~~~~$\hat{\bf d}(i)=\frac{1}{i}\sum\limits_{k=1}^i{\bf x}(k){y^*}(k)$ \\
~~~~{\bf Step 1. Compute the desired signal power} \\
~~~~$\hat{\sigma}^2_1(i)=\frac{|{{\hat{\bf a}_1}^H}(i){\bf x}(i)|^2-|{{\hat{\bf a}_1}^H}(i){\hat{\bf a}_1}(i)|\hat{\sigma}^2_n(i)}{|{{\hat{\bf a}_1}^H}(i){\hat{\bf a}_1}(i)|^2}$ \\
~~~~{\bf Step 2. Determine the Krylov subspace} \\
~~~~$\hat{\bf b}(i)=\hat{\bf a}_1(i)\hat{\bf a}_1^H(i)\hat{\bf a}_1(i)\hat{\sigma}_1^2(i)+\hat{\sigma}_n^2(i)\hat{\bf a}_1(i)$ \\
~~~~$\hat{\bf r}(i)=\hat{\bf b}(i)-\hat{\bf R}(i)\hat{\bf a}_1(i)$ \\
~~~~${\bf t}_1(i)=\frac{\hat{\bf r}(i)}{\lVert{\hat{\bf r}(i)}\rVert}$ \\
~~~~Apply the algorithm in Table \ref{table1} to determine $m$ and ${\bf t}_1(i)$,$\dotsb$,${\bf t}_m(i)$ \\
~~~~$\hat{\bf T}(i)=[{\bf t}_{1}(i), {\bf t}_{2}(i), \dotsb, {\bf t}_{m}(i)]$ \\
~~~~{\bf Step 3. Update the steering vector} \\
~~~~$\hat{\bf P}(i)=\hat{\bf T}(i)\hat{\bf T}^H(i)$ \\
~~~~$\hat{\bf a}_1(i+1)=\hat{\bf a}_1(i)+\frac{\hat{\bf P}(i)\hat{\bf d}(i)}{\lVert\hat{\bf P}(i)\hat{\bf d}(i)\rVert}$ \\
~~~~$\hat{\bf a}_1(i+1)=\hat{\bf a}_1(i+1)/\lVert\hat{\bf a}_1(i+1)\rVert$ \\
~~~~{\bf Step 4. Compute the weight vector} \\
~~~~$\hat{\bf R}_{\textcolor{red}{I+N}}(i)=\hat{\bf R}(i)-\hat{\sigma}^2_1(i)\hat{\bf a}_1(i)\hat{\bf a}^H_1(i)$ \\
~~~~$\hat{\bf w}(i)=\frac{{\hat{\bf R}^{-1}_{\textcolor{red}{I+N}}}(i)\hat{\bf a}_1(i)}{\hat{\bf a}^H_1(i)\hat{\bf R}^{-1}_{\textcolor{red}{I+N}}(i)\hat{\bf a}_1(i)}$ \\
End snapshot \\
\hline
\end{tabular} \label{table2}
\end{center}
\end{table}

\section{Proposed Adaptive Algorithms}

This section presents adaptive strategies based on the OKSPME robust beamforming method, resulting in the proposed OKSPME-SG, OKSPME-CCG and OKSPME-MCG algorithms, which are especially suitable for dynamic scenarios. In the proposed adaptive algorithms, we estimate the desired signal power and its steering vector with the same recursions as in OKSPME, whereas the estimation procedure of the beamforming weights is different. In particular, we start from a reformulated optimization problem and use SG and CG-based adaptive recursions to derive the weight update equations, which reduce the complexity by an order of magnitude as compared to that of OKSPME.

\subsection{{OKSPME-SG Adaptive Algorithm}}

We resort to an SG adaptive strategy and consider the following optimization problem:
\begin{equation}
\begin{aligned}
& \underset{{\bf w}(i)} {\text{minimize}}
&& {\bf w}^H(i)(\hat{\bf R}(i)-\hat{\bf R}_1(i)){\bf w}(i) \\
& \text{subject to} && {\bf w}^H(i)\hat{\bf a}_1(i)=1, \label{eq28}
\end{aligned}
\end{equation}
where $\hat{\bf R}(i)$ can be written as ${\bf x}(i){\bf x}^H(i)$ and $\hat{\bf R}_1(i)$ represents the desired signal covariance matrix and can be written as $\hat{\sigma}^2_1(i){\hat{\bf a}_1}(i){\hat{\bf a}^H_1}(i)$.

Then we can express the SG recursion as
\begin{equation}
{\bf w}(i+1)={\bf w}(i)-\mu\frac{\partial{\mathcal L}}{\partial{\bf w}(i)}, \label{eq29}
\end{equation}
where ${\mathcal L}={\bf w}^H(i)({\bf x}(i){\bf x}^H(i)-\hat{\sigma}^2_1(i){\hat{\bf a}_1}(i){\hat{\bf a}^H_1}(i)){\bf w}(i)+{\lambda}_{\mathcal L}({\bf w}^H(i){\hat{\bf a}_1}(i)-1)$ and $\mu$ is the step size.

By substituting ${\mathcal L}$ into the SG equation \eqref{eq29} and letting ${\bf w}^H(i+1)\hat{\bf a}_1(i+1)=1$, ${\lambda}_{\mathcal L}$ is obtained as
\begin{equation}
{\lambda}_{\mathcal L}=\frac{2(\hat{\sigma}^2_1(i){\hat{\bf a}^H_1}(i){\hat{\bf a}_1}(i)-y(i){\bf x}^H(i){\hat{\bf a}_1}(i))}{{\hat{\bf a}^H_1}(i){\hat{\bf a}_1}(i)}. \label{eq30}
\end{equation}

By substituting ${\lambda}_{\mathcal L}$ in \eqref{eq29} again, the weight update equation for OKSPME-SG is obtained as
\begin{equation}
\begin{aligned}
{\bf w}(i+1) & =({\bf I}-\mu\hat{\sigma}^2_1(i){\hat{\bf a}_1}(i){\hat{\bf a}^H_1}(i)){\bf w}(i) \\
& -\mu(\hat{\sigma}^2_1(i){\hat{\bf a}_1}(i)+y^*(i)({\bf x}(i)-\frac{\hat{\bf a}_1^H(i){\bf x}(i)\hat{\bf a}_1(i)}{\hat{\bf a}_1^H(i)\hat{\bf a}_1(i)})).
\label{eq31}
\end{aligned}
\end{equation}

The adaptive SG recursion circumvents a matrix inversion when computing the weights using \eqref{eq27}, which is unavoidable in OKSPME. Therefore, the computational complexity is reduced from ${\mathcal O}(M^3)$ in OKSPME to ${\mathcal O}(M^2)$ in OKSPME-SG. It is also important that the step size $\mu$ should satisfy $0<\mu<\frac{1}{\hat{\sigma}^2_1(i)}$ to guarantee that ${\bf I}-\mu\hat{\sigma}^2_1(i){\hat{\bf a}_1}(i){\hat{\bf a}^H_1}(i)$ is always a positive-definite matrix so that \eqref{eq31} is ensured converging to a solution. To implement OKSPME-SG, {\bf Step 1}, {\bf Step 2} and {\bf Step 3} from Table \ref{table2} and \eqref{eq31} are required.

\subsection{{OKSPME-CCG Adaptive Algorithm}}

In this subsection, the proposed OKSPME-CCG algorithm is introduced. In CG-based approaches, we usually employ a forgetting factor (e.g. $\lambda$) to estimate the second-order statistics of the data or the SCM \cite{r1,r10}, which can be expressed by
\begin{equation}
\hat{\bf R}(i)=\lambda\hat{\bf R}(i-1)+{\bf x}(i){\bf x}^H(i), \label{eq32}
\end{equation}
whereas the SCV $\hat{\bf d}(i)$ can be estimated with the same forgetting factor as described by
\begin{equation}
\hat{\bf d}(i)=\lambda\hat{\bf d}(i-1)+{\bf x}(i)y^*(i). \label{eq33}
\end{equation}

The proposed optimization problem that leads to the OKSPME-CCG algorithm is described by
\begin{equation}
\underset{\hat{\bf a}_1(i),{\bf v}(i)} {\text{minimize}}
{\mathcal J}={\bf v}^H(i)(\hat{\bf R}(i)-\hat{\bf R}_1(i)){\bf v}(i)-\hat{\bf a}_1^H(i){\bf v}(i), \label{eq34}
\end{equation}
where ${\bf v}(i)$ is the CG-based weight vector. In OKSPME-CCG, we require $N$ iterations for each snapshot. In the $n$th iteration, $\hat{\bf a}_{1,n}(i)$ and ${\bf v}_n(i)$ are updated as follows
\begin{equation}
\hat{\bf a}_{1,n}(i)=\hat{\bf a}_{1,n-1}(i)+{\alpha}_{\hat{\bf a}_1,n}(i){\bf p}_{\hat{\bf a}_1,n}(i), \label{eq35}
\end{equation}
\begin{equation}
{\bf v}_n(i)={\bf v}_{n-1}(i)+{\alpha}_{{\bf v},n}(i){\bf p}_{{\bf v},n}(i), \label{eq36}
\end{equation}
where ${\bf p}_{\hat{\bf a}_1,n}(i)$ and ${\bf p}_{{\bf v},n}(i)$ are direction vectors updated by
\begin{equation}
{\bf p}_{\hat{\bf a}_1,n+1}(i)={\bf g}_{\hat{\bf a}_1,n}(i)+{\beta}_{\hat{\bf a}_1,n}(i){\bf p}_{\hat{\bf a}_1,n}(i), \label{eq37}
\end{equation}
\begin{equation}
{\bf p}_{{\bf v},n+1}(i)={\bf g}_{{\bf v},n}(i)+{\beta}_{{\bf v},n}(i){\bf p}_{{\bf v},n}(i), \label{eq38}
\end{equation}
where ${\bf g}_{\hat{\bf a}_1,n}(i)$ and ${\bf g}_{{\bf v},n}(i)$ are the negative gradients of the cost function in terms of $\hat{\bf a}_1(i)$ and ${\bf v}(i)$, respectively, which are expressed as
\begin{equation}
{\bf g}_{\hat{\bf a}_1,n}(i)=-\frac{\partial{\mathcal J}}{\partial{\hat{\bf a}_{1,n}(i)}}=\hat{\sigma}^2_1(i){\bf v}_n(i){\bf v}^H_n(i)\hat{\bf a}_{1,n}(i)+{\bf v}_n(i), \label{eq39}
\end{equation}
\begin{multline}
{\bf g}_{{\bf v},n}(i)=-\frac{\partial{\mathcal J}}{\partial{{\bf v}_n(i)}} \\
={\bf g}_{{\bf v},n-1}(i)-{\alpha}_{{\bf v},n}(i)(\hat{\bf R}(i)-\hat{\sigma}^2_1(i){\bf x}(i){\bf x}^H(i)){\bf p}_{{\bf v},n}(i).  \label{eq40}
\end{multline}

The scaling parameters ${\alpha}_{\hat{\bf a}_1,n}(i)$, ${\alpha}_{{\bf v},n}(i)$ can be obtained by substituting \eqref{eq35} and \eqref{eq36} into \eqref{eq34} and minimizing the cost function with respect to ${\alpha}_{\hat{\bf a}_1,n}(i)$ and ${\alpha}_{{\bf v},n}(i)$, respectively. The solutions are given by
\begin{equation}
{\alpha}_{\hat{\bf a}_1,n}(i)=-\frac{{\bf g}^H_{\hat{\bf a}_1,n-1}(i){\bf p}_{\hat{\bf a}_1,n}(i)}{\hat{\sigma}^2_1(i){\bf p}^H_{\hat{\bf a}_1,n}(i){\bf v}_n(i){\bf v}^H_n(i){\bf p}_{\hat{\bf a}_1,n}(i)}, \label{eq41}
\end{equation}
\begin{equation}
{\alpha}_{{\bf v},n}(i)=\frac{{\bf g}^H_{{\bf v},n-1}(i){\bf p}_{{\bf v},n}(i)}{{\bf p}^H_{{\bf v},n}(i)(\hat{\bf R}(i)-\hat{\sigma}^2_1(i)\hat{\bf a}_{1,n}(i)\hat{\bf a}^H_{1,n}(i)){\bf p}_{{\bf v},n}(i)}. \label{eq42}
\end{equation}

The parameters ${\beta}_{\hat{\bf a}_1,n}(i)$ and ${\beta}_{{\bf v},n}(i)$ should be chosen to provide conjugacy for direction vectors \cite{r10}, which results in
\begin{equation}
{\beta}_{\hat{\bf a}_1,n}(i)=\frac{{\bf g}^H_{\hat{\bf a}_1,n}(i){\bf g}_{\hat{\bf a}_1,n}(i)}{{\bf g}^H_{\hat{\bf a}_1,n-1}(i){\bf g}_{\hat{\bf a}_1,n-1}(i)}, \label{eq43}
\end{equation}
\begin{equation}
{\beta}_{{\bf v},n}(i)=\frac{{\bf g}^H_{{\bf v},n}(i){\bf g}_{{\bf v},n}(i)}{{\bf g}^H_{{\bf v},n-1}(i){\bf g}_{{\bf v},n-1}(i)}. \label{eq44}
\end{equation}

After $\hat{\bf a}_{1,n}(i)$ and ${\bf v}_n(i)$ are updated for $N$ iterations, the beamforming weight vector ${\bf w}(i)$ can be computed by
\begin{equation}
{\bf w}(i)=\frac{{\bf v}_N(i)}{\hat{\bf a}^H_{1,N}(i){\bf v}_N(i)}, \label{eq45}
\end{equation}
The computational cost of OKSPME-CCG algorithm is ${\mathcal O}(NM^2)$, which is higher than the cost required in OKSPME-SG due to the inner iterations at every snapshot. The proposed OKSPME-CCG is summarized in Table \ref{table3}.

\begin{table}
\small
\begin{center}
\caption{Proposed OKSPME-CCG algorithm}
\begin{tabular}{l}
\hline
Initialization: \\
$\hat{\bf w}(1)={\bf v}_0(1)={\bf 1}$; $\lambda$; \\
Choose an initial guess $\hat{\bf a}_1(0)$ within the sector and set $\hat{\bf a}_1(1)=\hat{\bf a}_1(0)$; \\
For each snapshot $i=1,2,\dotsb$: \\
~~~~$\hat{\bf R}(i)=\frac{1}{i}\sum\limits_{k=1}^i{\bf x}(k){{\bf x}^H}(k)$ \\
~~~~$\hat{\bf d}(i)=\frac{1}{i}\sum\limits_{k=1}^i{\bf x}(k){y^*}(k)$ \\
~~~~{\bf Step 1} from Table \ref{table2} \\
~~~~{\bf Step 2} from Table \ref{table2} \\
~~~~{\bf Step 3} from Table \ref{table2} \\
~~~~\bf {Steering Vector and Weight Vector Estimations} \\
~~~~$\hat{\bf a}_{1,0}(i)=\hat{\bf a}_1(i)$ \\
~~~~${\bf g}_{\hat{\bf a}_1,0}(i)=\hat{\sigma}^2_1(i){\bf v}_0(i){\bf v}^H_0(i)\hat{\bf a}_{1,0}(i)+{\bf v}_0(i)$ \\
~~~~${\bf g}_{{\bf v},0}(i)=\hat{\bf a}_{1,0}(i)-\hat{\bf R}(i){\bf v}_0(i)$ \\
~~~~${\bf p}_{\hat{\bf a}_1,0}(i)={\bf g}_{\hat{\bf a}_1,0}(i)$; ${\bf p}_{{\bf v},0}(i)={\bf g}_{{\bf v},0}(i)$ \\
~~~~For each iteration index $n=1,2,\dotsb,N$: \\
~~~~~~~~${\alpha}_{\hat{\bf a}_1,n}(i)=-\frac{{\bf g}^H_{\hat{\bf a}_1,n-1}(i){\bf p}_{\hat{\bf a}_1,n}(i)}{\hat{\sigma}^2_1(i){\bf p}^H_{\hat{\bf a}_1,n}(i){\bf v}_n(i){\bf v}^H_n(i){\bf p}_{\hat{\bf a}_1,n}(i)}$ \\
~~~~~~~~${\alpha}_{{\bf v},n}(i)=\frac{{\bf g}^H_{{\bf v},n-1}(i){\bf p}_{{\bf v},n}(i)}{{\bf p}^H_{{\bf v},n}(i)(\hat{\bf R}(i)-\hat{\sigma}^2_1(i)\hat{\bf a}_{1,n}(i)\hat{\bf a}^H_{1,n}(i)){\bf p}_{{\bf v},n}(i)}$ \\
~~~~~~~~$\hat{\bf a}_{1,n}(i)=\hat{\bf a}_{1,n-1}(i)+{\alpha}_{\hat{\bf a}_1,n}(i){\bf p}_{\hat{\bf a}_1,n}(i)$ \\
~~~~~~~~${\bf v}_n(i)={\bf v}_{n-1}(i)+{\alpha}_{{\bf v},n}(i){\bf p}_{{\bf v},n}(i)$ \\
~~~~~~~~${\bf g}_{\hat{\bf a}_1,n}(i)=\hat{\sigma}^2_1(i){\bf v}_n(i){\bf v}^H_n(i)\hat{\bf a}_{1,n}(i)+{\bf v}_n(i)$ \\
~~~~~~~~${\bf g}_{{\bf v},n}(i)={\bf g}_{{\bf v},n-1}(i)-{\alpha}_{{\bf v},n}(i)(\hat{\bf R}(i)-\hat{\sigma}^2_1(i){\bf x}(i){\bf x}^H(i)){\bf p}_{{\bf v},n}(i)$ \\
~~~~~~~~${\beta}_{\hat{\bf a}_1,n}(i)=\frac{{\bf g}^H_{\hat{\bf a}_1,n}(i){\bf g}_{\hat{\bf a}_1,n}(i)}{{\bf g}^H_{\hat{\bf a}_1,n-1}(i){\bf g}_{\hat{\bf a}_1,n-1}(i)}$ \\
~~~~~~~~${\beta}_{{\bf v},n}(i)=\frac{{\bf g}^H_{{\bf v},n}(i){\bf g}_{{\bf v},n}(i)}{{\bf g}^H_{{\bf v},n-1}(i){\bf g}_{{\bf v},n-1}(i)}$ \\
~~~~~~~~${\bf p}_{\hat{\bf a}_1,n+1}(i)={\bf g}_{\hat{\bf a}_1,n}(i)+{\beta}_{\hat{\bf a}_1,n}(i){\bf p}_{\hat{\bf a}_1,n}(i)$ \\
~~~~~~~~${\bf p}_{{\bf v},n+1}(i)={\bf g}_{{\bf v},n}(i)+{\beta}_{{\bf v},n}(i){\bf p}_{{\bf v},n}(i)$ \\
~~~~End iteration \\
~~~~${\bf v}_0(i+1)={\bf v}_N(i)$ \\
~~~~${\bf w}(i)=\frac{{\bf v}_N(i)}{\hat{\bf a}^H_{1,N}(i){\bf v}_N(i)}$ \\
End snapshot \\
\hline
\end{tabular} \label{table3}
\end{center}
\end{table}

\subsection{{OKSPME-MCG Adaptive Algorithm}}

In OKSPME-MCG, we let only one iteration be performed per snapshot, which further reduces the complexity compared to OKSPME-CCG. Here we denote the CG-based weights and steering vector updated by snapshots rather than inner iterations as
\begin{equation}
\hat{\bf a}_1(i)=\hat{\bf a}_1(i-1)+{\alpha}_{\hat{\bf a}_1}(i){\bf p}_{\hat{\bf a}_1}(i), \label{eq46}
\end{equation}
\begin{equation}
{\bf v}(i)={\bf v}(i-1)+{\alpha}_{\bf v}(i){\bf p}_{\bf v}(i). \label{eq47}
\end{equation}

As can be seen, the subscripts of all the quantities for inner iterations are eliminated. Then, we employ the degenerated scheme to ensure ${\alpha}_{\hat{\bf a}_1}(i)$ and ${\alpha}_{\bf v}(i)$ satisfy the convergence bound \cite{r10} given by
\begin{equation}
0\leq{\bf p}^H_{\hat{\bf a}_1}(i){\bf g}_{\hat{\bf a}_1}(i)\leq0.5{\bf p}^H_{\hat{\bf a}_1}(i){\bf g}_{\hat{\bf a}_1}(i-1), \label{eq48}
\end{equation}
\begin{equation}
0\leq{\bf p}^H_{\bf v}(i){\bf g}_{\bf v}(i)\leq0.5{\bf p}^H_{\bf v}(i){\bf g}_{\bf v}(i-1). \label{eq49}
\end{equation}

Instead of updating the negative gradient vectors ${\bf g}_{\hat{\bf a}_1}(i)$ and ${\bf g}_{\bf v}(i)$ in iterations, now we utilize the forgetting factor to re-express them in one snapshot as
\begin{equation}
\begin{aligned}
{\bf g}_{\hat{\bf a}_1}(i)=
& (1-\lambda){\bf v}(i)+\lambda{\bf g}_{\hat{\bf a}_1}(i-1) \\
& +\hat{\sigma}^2_1(i){\alpha}_{\hat{\bf a}_1}(i){\bf v}(i){\bf v}^H(i){\bf p}_{\hat{\bf a}_1}(i) \\
& -{\bf x}(i){\bf x}^H(i)\hat{\bf a}_1(i), \label{eq50}
\end{aligned}
\end{equation}
\begin{equation}
\begin{aligned}
{\bf g}_{\bf v}(i)=
& (1-\lambda)\hat{\bf a}_1(i)+\lambda{\bf g}_{\bf v}(i-1) \\
& -{\alpha}_{\bf v}(i)(\hat{\bf R}(i)-\hat{\sigma}^2_1(i)\hat{\bf a}_1(i)\hat{\bf a}^H_1(i)){\bf p}_{\bf v}(i) \\
& -{\bf x}(i){\bf x}^H(i){\bf v}(i-1). \label{eq51}
\end{aligned}
\end{equation}

Pre-multiplying \eqref{eq50} and \eqref{eq51} by ${\bf p}^H_{\hat{\bf a}_1}(i)$ and ${\bf p}^H_{\bf v}(i)$, respectively, and taking expectations we obtain
\begin{multline}
E[{\bf p}^H_{\hat{\bf a}_1}(i){\bf g}_{\hat{\bf a}_1}(i)]=E[{\bf p}^H_{\hat{\bf a}_1}(i)({\bf v}(i)-{\bf x}(i){\bf x}^H(i)\hat{\bf a}_1)(i)] \\
+{\lambda}E[{\bf p}^H_{\hat{\bf a}_1}(i){\bf g}_{\hat{\bf a}_1}(i-1)]-{\lambda}E[{\bf p}^H_{\hat{\bf a}_1}(i){\bf v}(i)] \\
+E[{\alpha}_{\hat{\bf a}_1}(i){\bf p}^H_{\hat{\bf a}_1}(i)\hat{\sigma}^2_1(i){\bf v}(i){\bf v}^H(i){\bf p}_{\hat{\bf a}_1}(i)], \label{eq52}
\end{multline}
\begin{multline}
E[{\bf p}^H_{\bf v}(i){\bf g}_{\bf v}(i)]={\lambda}E[{\bf p}^H_{\bf v}(i){\bf g}_{\bf v}(i-1)]-{\lambda}E[{\bf p}^H_{\bf v}(i)\hat{\bf a}_1(i)] \\
-E[{\alpha}_{\bf v}(i){\bf p}^H_{\bf v}(i)(\hat{\bf R}(i)-\hat{\sigma}^2_1(i)\hat{\bf a}_1(i)\hat{\bf a}^H_1(i)){\bf p}_{\bf v}(i)], \label{eq53}
\end{multline}
where in \eqref{eq53} we have $E[\hat{\bf R}(i){\bf v}(i-1)]=E[\hat{\bf a}_1(i)]$. After substituting \eqref{eq53} in \eqref{eq49} we obtain the bounds for ${\alpha}_{\bf v}(i)$ as follows
\begin{multline}
\frac{(\lambda-0.5)E[{\bf p}^H_{\bf v}(i){\bf g}_{\bf v}(i-1)]-{\lambda}E[{\bf p}^H_{\bf v}(i)\hat{\bf a}_1(i)]}{E[{\bf p}^H_{\bf v}(i)(\hat{\bf R}(i)-\hat{\sigma}^2_1(i)\hat{\bf a}_1(i)\hat{\bf a}^H_1(i)){\bf p}_{\bf v}(i)]}{\leq}E[{\alpha}_{\bf v}(i)] \\
{\leq}\frac{{\lambda}E[{\bf p}^H_{\bf v}(i){\bf g}_{\bf v}(i-1)]-{\lambda}E[{\bf p}^H_{\bf v}(i)\hat{\bf a}_1(i)]}{E[{\bf p}^H_{\bf v}(i)(\hat{\bf R}(i)-\hat{\sigma}^2_1(i)\hat{\bf a}_1(i)\hat{\bf a}^H_1(i)){\bf p}_{\bf v}(i)]}. \label{eq54}
\end{multline}

Then we can introduce a constant parameter ${\eta}_{\bf v} \in [0,0.5]$ to restrict ${\alpha}_{\bf v}(i)$ within the bounds in \eqref{eq54} as
\begin{multline}
{\alpha}_{\bf v}(i)= \\
\frac{\lambda({\bf p}^H_{\bf v}(i){\bf g}_{\bf v}(i-1)-{\bf p}^H_{\bf v}(i)\hat{\bf a}_1(i))-{\eta}_{\bf v}{\bf p}^H_{\bf v}(i){\bf g}_{\bf v}(i-1)}{{\bf p}^H_{\bf v}(i)(\hat{\bf R}(i)-\hat{\sigma}^2_1(i)\hat{\bf a}_1(i)\hat{\bf a}^H_1(i)){\bf p}_{\bf v}(i)}. \label{eq55}
\end{multline}

Similarly, we can also obtain the bounds for ${\alpha}_{\hat{\bf a}_1}(i)$. For simplicity let us define $E[{\bf p}^H_{\hat{\bf a}_1}(i){\bf g}_{\hat{\bf a}_1}(i-1)]=A$, $E[{\bf p}^H_{\hat{\bf a}_1}(i){\bf v}(i)]=B$, $E[{\bf p}^H_{\hat{\bf a}_1}(i){\bf x}(i){\bf x}^H(i)\hat{\bf a}_1(i)]=C$ and $E[{\bf p}^H_{\hat{\bf a}_1}(i)\hat{\sigma}^2_1(i){\bf v}(i){\bf v}^H(i){\bf p}_{\hat{\bf a}_1}(i)]=D$. Substituting \eqref{eq52} into \eqref{eq48} gives
\begin{multline}
\frac{\lambda(B-A)-B+C}{D}{\leq}E[{\alpha}_{\hat{\bf a}_1}(i)] \\
{\leq}\frac{\lambda(B-A)-B+C+0.5A}{D}, \label{eq56}
\end{multline}
in which we can introduce another constant parameter ${\eta}_{\hat{\bf a}_1} \in [0,0.5]$ to restrict ${\alpha}_{\hat{\bf a}_1}(i)$ within the bounds in \eqref{eq56} as
\begin{equation}
E[{\alpha}_{\hat{\bf a}_1}(i)]=\frac{\lambda(B-A)-B+C+{\eta}_{\hat{\bf a}_1}A}{D}, \label{eq57}
\end{equation}
or
\begin{multline}
{\alpha}_{\hat{\bf a}_1}(i)=[\lambda({\bf p}^H_{\hat{\bf a}_1}(i){\bf v}(i)-{\bf p}^H_{\hat{\bf a}_1}(i){\bf g}_{\hat{\bf a}_1}(i-1))-{\bf p}^H_{\hat{\bf a}_1}(i){\bf v}(i) \\
+{\bf p}^H_{\hat{\bf a}_1}(i){\bf x}(i){\bf x}^H(i)\hat{\bf a}_1(i)+{\eta}_{\hat{\bf a}_1}{\bf p}^H_{\hat{\bf a}_1}(i){\bf g}_{\hat{\bf a}_1}(i-1)] \\
/[\hat{\sigma}^2_1(i){\bf p}^H_{\hat{\bf a}_1}(i){\bf v}(i){\bf v}^H(i){\bf p}_{\hat{\bf a}_1}(i)]. \label{eq58}
\end{multline}

Then we can update the direction vectors ${\bf p}_{\hat{\bf a}_1}(i)$ and ${\bf p}_{\bf v}(i)$ by
\begin{equation}
{\bf p}_{\hat{\bf a}_1}(i+1)={\bf g}_{\hat{\bf a}_1}(i)+{\beta}_{\hat{\bf a}_1}(i){\bf p}_{\hat{\bf a}_1}(i), \label{eq59}
\end{equation}
\begin{equation}
{\bf p}_{\bf v}(i+1)={\bf g}_{\bf v}(i)+{\beta}_{\bf v}(i){\bf p}_{\bf v}(i), \label{eq60}
\end{equation}
where ${\beta}_{\hat{\bf a}_1}(i)$ and ${\beta}_{\bf v}(i)$ are updated by
\begin{equation}
{\beta}_{\hat{\bf a}_1}(i)=\frac{[{\bf g}_{\hat{\bf a}_1}(i)-{\bf g}_{\hat{\bf a}_1}(i-1)]^H{\bf g}_{\hat{\bf a}_1}(i)}{{\bf g}^H_{\hat{\bf a}_1}(i-1){\bf g}_{\hat{\bf a}_1}(i-1)}, \label{eq61}
\end{equation}
\begin{equation}
{\beta}_{\bf v}(i)=\frac{[{\bf g}_{\bf v}(i)-{\bf g}_{\bf v}(i-1)]^H{\bf g}_{\bf v}(i)}{{\bf g}^H_{\bf v}(i-1){\bf g}_{\bf v}(i-1)}. \label{eq62}
\end{equation}

Finally we can update the beamforming weights by
\begin{equation}
{\bf w}(i)=\frac{{\bf v}(i)}{\hat{\bf a}^H_1(i){\bf v}(i)}, \label{eq63}
\end{equation}

The MCG approach employs the forgetting factor $\lambda$ and constant $\eta$ for estimating $\alpha(i)$, which means its performance may depend on a suitable choice of these parameters. The proposed OKSPME-MCG algorithm requires a complexity of ${\mathcal O}(M^2)$. However, the cost is usually much lower compared to CCG approach for the elimination of inner recursions and it presents a similar performance in most studied scenarios. From an implementation point of view, the choice of using the CCG and MCG algorithms is based on the stationarity of the system: the CCG algorithm is more suitable for scenarios in which the system is stationary and we can compute the beamformer with K iterations while the MCG algorithm is suggested for non-stationary scenarios as we only run one iteration per snapshot and can track variations in the environment. Table \ref{table4} summarizes the OKSPME-MCG algorithm.

\begin{table}
\small
\begin{center}
\caption{Proposed OKSPME-MCG algorithm}
\begin{tabular}{l}
\hline
Initialization: \\
$\hat{\bf w}(1)={\bf v}(0)={\bf 1}$; $\lambda$; ${\eta}_{\bf v}={\eta}_{\hat{\bf a}_1}$; \\
Choose an initial guess $\hat{\bf a}_1(0)$ within the sector and set $\hat{\bf a}_1(1)=\hat{\bf a}_1(0)$; \\
${\bf g}_{\bf v}(0)={\bf p}_{\bf v}(1)=\hat{\bf a}_1(1)$;
${\bf g}_{\hat{\bf a}_1}(0)={\bf p}_{\hat{\bf a}_1}(1)={\bf v}(0)$; \\
For each snapshot $i=1,2,\dotsb$: \\
~~~~$\hat{\bf R}(i)=\frac{1}{i}\sum\limits_{k=1}^i{\bf x}(k){{\bf x}^H}(k)$ \\
~~~~$\hat{\bf d}(i)=\frac{1}{i}\sum\limits_{k=1}^i{\bf x}(k){y^*}(k)$ \\
~~~~{\bf Step 1} from Table \ref{table2} \\
~~~~{\bf Step 2} from Table \ref{table2} \\
~~~~{\bf Step 3} from Table \ref{table2} \\
~~~~\bf {Steering Vector and Weight Vector Estimations} \\
~~~~${\alpha}_{\hat{\bf a}_1}(i)=[\lambda({\bf p}^H_{\hat{\bf a}_1}(i){\bf v}(i)-{\bf p}^H_{\hat{\bf a}_1}(i){\bf g}_{\hat{\bf a}_1}(i-1))-{\bf p}^H_{\hat{\bf a}_1}(i){\bf v}(i)$ \\
~~~~$+{\bf p}^H_{\hat{\bf a}_1}(i){\bf x}(i){\bf x}^H(i)\hat{\bf a}_1(i)+{\eta}_{\hat{\bf a}_1}{\bf p}^H_{\hat{\bf a}_1}(i){\bf g}_{\hat{\bf a}_1}(i-1)]$ \\
~~~~$/[\hat{\sigma}^2_1(i){\bf p}^H_{\hat{\bf a}_1}(i){\bf v}(i){\bf v}^H(i){\bf p}_{\hat{\bf a}_1}(i)]$ \\
~~~~${\alpha}_{\bf v}(i)=
\frac{\lambda({\bf p}^H_{\bf v}(i){\bf g}_{\bf v}(i-1)-{\bf p}^H_{\bf v}(i)\hat{\bf a}_1(i))-{\eta}_{\bf v}{\bf p}^H_{\bf v}(i){\bf g}_{\bf v}(i-1)}{{\bf p}^H_{\bf v}(i)(\hat{\bf R}(i)-\hat{\sigma}^2_1(i)\hat{\bf a}_1(i)\hat{\bf a}^H_1(i)){\bf p}_{\bf v}(i)}$ \\
~~~~$\hat{\bf a}_1(i)=\hat{\bf a}_1(i-1)+{\alpha}_{\hat{\bf a}_1}(i){\bf p}_{\hat{\bf a}_1}(i)$ \\
~~~~${\bf v}(i)={\bf v}(i-1)+{\alpha}_{\bf v}(i){\bf p}_{\bf v}(i)$ \\
~~~~${\bf g}_{\hat{\bf a}_1}(i)=(1-\lambda){\bf v}(i)+\lambda{\bf g}_{\hat{\bf a}_1}(i-1)$ \\
~~~~$+\hat{\sigma}^2_1(i){\alpha}_{\hat{\bf a}_1}(i){\bf v}(i){\bf v}^H(i){\bf p}_{\hat{\bf a}_1}(i)-{\bf x}(i){\bf x}^H(i)\hat{\bf a}_1(i)$ \\
~~~~${\bf g}_{\bf v}(i)=(1-\lambda)\hat{\bf a}_1(i)+\lambda{\bf g}_{\bf v}(i-1)-{\alpha}_{\bf v}(i)(\hat{\bf R}(i)$ \\
~~~~$-\hat{\sigma}^2_1(i)\hat{\bf a}_1(i)\hat{\bf a}^H_1(i)){\bf p}_{\bf v}(i)-{\bf x}(i){\bf x}^H(i){\bf v}(i-1)$ \\
~~~~${\beta}_{\hat{\bf a}_1}(i)=\frac{[{\bf g}_{\hat{\bf a}_1}(i)-{\bf g}_{\hat{\bf a}_1}(i-1)]^H{\bf g}_{\hat{\bf a}_1}(i)}{{\bf g}^H_{\hat{\bf a}_1}(i-1){\bf g}_{\hat{\bf a}_1}(i-1)}$ \\
~~~~${\beta}_{\bf v}(i)=\frac{[{\bf g}_{\bf v}(i)-{\bf g}_{\bf v}(i-1)]^H{\bf g}_{\bf v}(i)}{{\bf g}^H_{\bf v}(i-1){\bf g}_{\bf v}(i-1)}$ \\
~~~~${\bf p}_{\hat{\bf a}_1}(i+1)={\bf g}_{\hat{\bf a}_1}(i)+{\beta}_{\hat{\bf a}_1}(i){\bf p}_{\hat{\bf a}_1}(i)$ \\
~~~~${\bf p}_{\bf v}(i+1)={\bf g}_{\bf v}(i)+{\beta}_{\bf v}(i){\bf p}_{\bf v}(i)$ \\
~~~~${\bf w}(i)=\frac{{\bf v}(i)}{\hat{\bf a}^H_1(i){\bf v}(i)}$ \\
End snapshot \\
\hline
\end{tabular} \label{table4}
\end{center}
\end{table}

\section{Analysis}

In this section, we present an analysis of the following aspects of the proposed and existing algorithms:
\begin{itemize}
\item An analysis of the MSE between the estimated and actual steering vectors for the general approach that employs a presumed angular sector.
\item MSE analysis results of the proposed OKSPME method and the SQP method in \cite{r6} and their relationships and differences.
\item A complexity analysis for the proposed and existing algorithms.
\end{itemize}

\subsection{MSE analysis}

\begin{figure}[!htb]
\begin{center}
\def\epsfsize#1#2{0.6\columnwidth}
\epsfbox{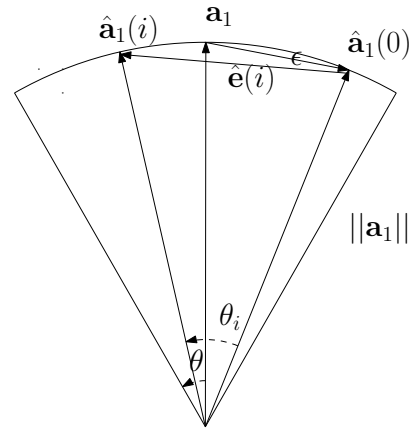}
\caption{Euclidean norm interpretation of the MSE} \label{ipe1}
\end{center}
\end{figure}

Firstly, we present the MSE analysis of the general approach that employs a presumed angular sector. Since we have the steering vector estimate $\hat{\bf a}_1(i)$ in the $i$th snapshot, by denoting the true steering vector as ${\bf a}_1$, we can express the MSE of the estimate $\hat{\bf a}_1(i)$ as
\begin{multline}
{\rm MSE}\{\hat{\bf a}_1(i)\}={\rm tr}(E[(\hat{\bf a}_1(i)-{\bf a}_1)(\hat{\bf a}_1(i)-{\bf a}_1)^H]) \\
=E[(\hat{\bf a}_1(i)-{\bf a}_1)^H(\hat{\bf a}_1(i)-{\bf a}_1)]. \label{eq75}
\end{multline}
In the approach that employs an angular sector, we usually choose an initial guess (i.e., $\hat{\bf a}_1(0)$) from the presumed sector. Let us express the accumulated estimation error as
\begin{equation}
\hat{\bf e}(i)=\hat{\bf a}_1(i)-\hat{\bf a}_1(0), \label{eq76}
\end{equation}
then \eqref{eq75} can be rewritten as
\begin{equation}
{\rm MSE}\{\hat{\bf a}_1(i)\} \\
=E[(\hat{\bf a}_1(0)+\hat{\bf e}(i)-{\bf a}_1)^H(\hat{\bf a}_1(0)+\hat{\bf e}(i)-{\bf a}_1)]. \label{eq77}
\end{equation}
The initial guess $\hat{\bf a}_1(0)$ can be described as the true steering vector plus a guess error vector (i.e., $\bf\epsilon$):
\begin{equation}
\hat{\bf a}_1(0)={\bf a}_1+\bf\epsilon. \label{eq78}
\end{equation}
Taking expectation of both sides of the above, we have
\begin{equation}
E[\hat{\bf a}_1(0)]={\bf a}_1+E[\bf\epsilon]. \label{eq79}
\end{equation}
Substituting \eqref{eq78} into \eqref{eq77}, taking into account that the accumulated estimation error is uncorrelated with the initial guess error and simplifying the expression, we obtain
\begin{multline}
{\rm MSE}\{\hat{\bf a}_1(i)\} \\
=E[{\bf\epsilon}^H{\bf\epsilon}]+E[{\bf\epsilon}^H]E[\hat{\bf e}(i)]+E[\hat{\bf e}^H(i)]E[{\bf\epsilon}]+E[\hat{\bf e}^H(i)\hat{\bf e}(i)]. \label{eq80}
\end{multline}
Furthermore, it should be emphasized that both ${\bf\epsilon}$ and $\hat{\bf e}(i)$ are in vector forms, which means that their second-order statistics can be re-expressed in terms of their first-order statistics of their Euclidean norms. Then we can re-express \eqref{eq80} as
\begin{multline}
{\rm MSE}\{\hat{\bf a}_1(i)\} \\
=E[\lVert{\bf\epsilon}\rVert^2]+E[\lVert\hat{\bf e}(i)\rVert^2]+2{E[{\bf\epsilon}^H]E[\hat{\bf e}(i)]}. \label{eq81}
\end{multline}
Since both $\lVert{\bf\epsilon}\rVert$ and $\lVert\hat{\bf e}(i)\rVert$ are scalars we have
\begin{equation}
E[\lVert{\bf\epsilon}\rVert^2]={\rm Var}[\lVert{\bf\epsilon}\rVert]+E^2[\lVert{\bf\epsilon}\rVert], \label{eq82}
\end{equation}
\begin{equation}
E[\lVert\hat{\bf e}(i)\rVert^2]={\rm Var}[\lVert\hat{\bf e}(i)\rVert]+E^2[\lVert\hat{\bf e}(i)\rVert]. \label{eq83}
\end{equation}
At this stage, we can employ Popoviciu's inequality \cite{r37} to obtain the upper bounds for \textcolor{red}{the variances of the norms of the random vectors $\bf\epsilon$ and $\hat{\bf e}(i)$}, which are given by
\begin{equation}
{\rm Var}[\lVert{\bf\epsilon}\rVert] \leq \frac{(\sup\lVert{\bf\epsilon}\rVert-\inf\lVert{\bf\epsilon}\rVert)^2}{4}, \label{eq84}
\end{equation}
\begin{equation}
{\rm Var}[\lVert\hat{\bf e}(i)\rVert] \leq \frac{(\sup\lVert\hat{\bf e}(i)\rVert-\inf\lVert\hat{\bf e}(i)\rVert)^2}{4}.
\label{eq85}
\end{equation}
However, the last term in \eqref{eq81} is not analytical when conducting a norm analysis. Actually, $E[{\bf\epsilon}]$ depends on how the presumed sector is chosen: if the sector is chosen in an unbiased manner (i.e., the true steering vector lies in the centre of the sector), then we have $E[{\bf\epsilon}]={\bf 0}$ by symmetry criterion, in which case we can omit the last terms of $\eqref{eq81}$. For convenience of carrying out the norm analysis as the next step, we focus on the unbiased case only, so that the MSE only depends on the expectation, the infimum and the supremum of $\lVert{\bf\epsilon}\rVert$ and $\lVert\hat{\bf e}(i)\rVert$. In Fig. \ref{ipe1}, we utilize Euclidean geometry to illustrate the relationships among the norms of the errors and the norm of the steering vector, which is a fixed parameter due to the re-normalization procedure after it is estimated each time.

According to Fig. \ref{ipe1}, we can use $\theta$ (i.e., half of the angular sector, assumed less than $\pi/4$) and $\lVert{\bf a}_1\rVert$ to obtain $E[\lVert{\bf\epsilon}\rVert]$ by the following (any angular parameter appeared in the equations should be measured in radians rather than degrees): $\lVert{\bf\epsilon}\rVert$ is equivalent to the chord length which corresponds to the arc of a variable $\tau$, which can be any value from $0$ to $\theta$ with equal probability, in other words, the choice of $\tau$ is uniformly distributed within $[0,\theta]$. If the sample size of the selected $\bf\epsilon$ is large enough, we can approximately describe its probability density function (pdf) as a continuous function given by
\begin{equation}
{\rm f}(\tau)=\frac{1}{\theta}. \label{eq86}
\end{equation}
Meanwhile, we are also able to calculate the chord length $\lVert{\bf\epsilon}\rVert$ from a simple geometric criterion as
\begin{equation}
\lVert{\bf\epsilon}\rVert=2\lVert{\bf a}_1\rVert\sin\frac{\tau}{2}. \label{eq87}
\end{equation}
Then the expectation of $\lVert{\bf\epsilon}\rVert$ can be computed by
\begin{equation}
E[\lVert{\bf\epsilon}\rVert]=\int\limits_0^{\theta}\lVert{\bf\epsilon}\rVert{\rm f}(\tau){\rm d}\tau, \label{eq88}
\end{equation}
from which after simplification we obtain
\begin{equation}
E[\lVert{\bf\epsilon}\rVert]=\frac{8\lVert{\bf a}_1\rVert\sin^2\frac{\theta}{4}}{\theta}. \label{eq89}
\end{equation}
At this point, we can also compute the variance of $\lVert{\bf\epsilon}\rVert$ by using \eqref{eq89} as
\begin{equation}
{\rm Var}[\lVert{\bf\epsilon}\rVert]=\int\limits_0^{\theta}(\lVert{\bf\epsilon}\rVert-E[\lVert{\bf\epsilon}\rVert])^2{\rm f}(\tau){\rm d}\tau, \label{eq90}
\end{equation}
from which after simplification we obtain
\begin{equation}
{\rm Var}[\lVert{\bf\epsilon}\rVert]=2\lVert{\bf a}_1\rVert^2(1-\frac{\sin\theta}{\theta}-\frac{32\sin^4\frac{\theta}{4}}{\theta^2}). \label{eq91}
\end{equation}
In addition, it is clear that we have $\inf\lVert{\bf\epsilon}\rVert=0$ and $\sup\lVert{\bf\epsilon}\rVert=2\lVert{\bf a}_1\rVert\sin\frac{\theta}{2}$, which can be substituted in \eqref{eq84} and result in
\begin{equation}
{\rm Var}[\lVert{\bf\epsilon}\rVert] \leq \lVert{\bf a}_1\rVert^2\sin^2\frac{\theta}{2}. \label{eq92}
\end{equation}
We can see that the right-hand side of \eqref{eq91} satisfies the inequality in \eqref{eq92}. After substituting \eqref{eq89} and \eqref{eq91} in \eqref{eq82}, we obtain
\begin{equation}
E[\lVert{\bf\epsilon}\rVert^2]=2\lVert{\bf a}_1\rVert^2(1-\frac{\sin\theta}{\theta}). \label{eq93}
\end{equation}

\begin{figure}[!htb]
\begin{center}
\def\epsfsize#1#2{0.7\columnwidth}
\epsfbox{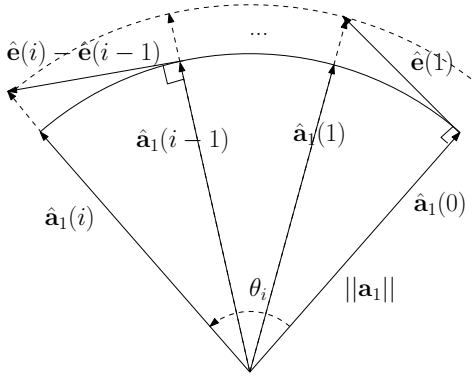}
\caption{update scheme of the SQP method} \label{ipe2}
\end{center}
\end{figure}

\begin{figure}[!htb]
\begin{center}
\def\epsfsize#1#2{0.8\columnwidth}
\epsfbox{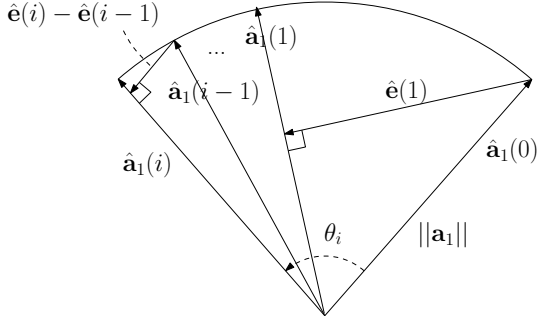}
\caption{update scheme of the OKSPME method} \label{ipe3}
\end{center}
\end{figure}

Regarding the computation of the norm of the accumulated estimation error $\lVert\hat{\bf e}(i)\rVert$, we need to emphasize that even though the steering vector is always re-normalized each time after it is updated, the piecewise estimation error in each snapshot does not directly update the steering vector to its normalized version, which means it is inappropriate to calculate the estimation error by geometric methods directly from Fig. \ref{ipe1} because the accumulated estimation error partially comes from the unnormalized steering vectors. However, we can obtain the infimum and supremum values for $\lVert\hat{\bf e}(i)\rVert$ if we assume the update scheme is unidirectional (i.e., the steering vector is updated from $\hat{\bf a}_1(0)$ to $\hat{\bf a}_1(i)$ in a single direction within the sector), with the unnormalized steering vectors considered.

We firstly look at the SQP method scenario in \cite{r6}. The steering vector update scheme is shown in Fig. \ref{ipe2}. It is necessary to emphasize that now we focus on the angular sector range of ${\theta}_i$ (i.e., the angle difference between the initially guessed steering vector and its estimate in the $i$th snapshot) rather than $\theta$. In \cite{r6}, an online-optimization program was used to iteratively solve for the piecewise estimation error in every snapshot, which was always orthogonal to the current steering vector estimate. Let us consider that at each time instant the steering vector is updated, its direction changes by ${\theta}_{i,k}$, where $i$ is the snapshot index and $k$ ($1\leq{k}\leq{i}$) is the index for the $k$th update. Since the total direction change in a snapshot is ${\theta}_i$, then we have
\begin{equation}
{\theta}_i=\sum\limits_{k=1}^i{\theta}_{i,k}, \label{eq94}
\end{equation}
and the norm of the accumulated estimation error is no greater than the sum of the norms of all the piecewise estimation errors, which is given by the inequality
\begin{equation}
\lVert\hat{\bf e}(i)\rVert \leq \sum\limits_{k=1}^{i}\lVert{\bf a}_1\rVert\tan{\theta}_{i,k}. \label{eq95}
\end{equation}
If we assume ${\theta}_i$ is less than $\pi/2$, then the right-hand side of \eqref{eq95} achieves its maximum value when ${\theta}_{i,k}=\tan{\theta}_i$, which is also the supremum of $\lVert\hat{\bf e}(i)\rVert$ and equals
\begin{equation}
{\lVert\hat{\bf e}(i)\rVert}_{max}=\lVert{\bf a}_1\rVert\tan{\theta}_i. \label{eq96}
\end{equation}
On the other hand, we notice that the piecewise estimation error vector can never enter into the angular sector, but at most move along with the arc if the number of iterations is large enough. In this case, we can approximately and geometrically illustrate the arc length corresponding with $\theta_{i}$ as the lower bound by taking the limit $i\rightarrow\infty$, i.e.,
\begin{equation}
\underset{i\rightarrow\infty}\lim{\lVert\hat{\bf e}(i)\rVert}={\theta}_i\lVert{\bf a}_1\rVert, \label{eq97}
\end{equation}
which is actually the infimum of $\lVert\hat{\bf e}(i)\rVert$ and cannot be achieved since the number of snapshots or iterations are always limited in practical situations. By combining \eqref{eq96} and \eqref{eq97}, ${\lVert\hat{\bf e}(i)\rVert}$ is bounded by
\begin{equation}
\inf\lVert\hat{\bf e}(i)\rVert = {\theta}_i\lVert{\bf a}_1\rVert < \lVert\hat{\bf e}(i)\rVert \leq \lVert{\bf a}_1\rVert\tan{\theta}_i = \sup\lVert\hat{\bf e}(i)\rVert. \label{eq98}
\end{equation}

Different from the SQP method, the proposed OKSPME method utilizes the Krylov subspace and the cross-correlation vector projection approach to extract the error information then use it to update the steering vector. From \eqref{eq9} we have
\begin{multline}
\hat{\bf d}(i)=\frac{1}{i}\sum\limits_{k=1}^i{\bf x}(k){y^*}(k)=\frac{1}{i}\sum\limits_{k=1}^i{\bf x}(k)({\bf w}^H(k){\bf x}(k))^* \\
=\frac{1}{i}\sum\limits_{k=1}^i{\bf x}(k){\bf x}^H(k){\bf w}(k)=\frac{1}{i}\sum\limits_{k=1}^i\hat{\bf R}(k){\bf w}(k).  \label{eq99}
\end{multline}
\textcolor{red}{Note that an initialization for vector $\hat{\bf d}$ or matrix $\hat{\bf R}$ should be considered to ensure $\hat{\bf R}$ is full-rank and invertible, which can be done by either setting $\hat{\bf d}(0)=\delta{\bf I}{\bf w}(0)$ or $\hat{\bf R}(0)=\delta{\bf I}$.}
We also know that
\begin{equation}
{\bf w}(k)=\frac{\hat{\bf R}^{-1}(k)\hat{\bf a}_1(k)}{{\hat{\bf a}_1}^H(k){\hat{\bf R}^{-1}(k)}\hat{\bf a}_1(k)}=\frac{\hat{\bf R}^{-1}(k)\hat{\bf a}_1(k)}{\hat{\sigma}^2_1(k)}. \label{eq100}
\end{equation}
Pre-multiplying \eqref{eq100} by $\hat{\bf R}(k)$ on both sides we obtain
\begin{equation}
\hat{\bf R}(k){\bf w}(k)=\frac{\hat{\bf a}_1(k)}{\hat{\sigma}^2_1(k)}, \label{eq101}
\end{equation}
which is then substituted in \eqref{eq99} and results in
\begin{equation}
\hat{\bf d}(i)=\frac{1}{i}\sum\limits_{k=1}^i\frac{\hat{\bf a}_1(k)}{\hat{\sigma}^2_1(k)}, \label{eq102}
\end{equation}
where $\hat{\sigma}^2_1(k)$ is a scalar, which means the SCV contains the direction of the desired signal steering vector.  Projecting $\hat{\bf d}(i)$ onto the Krylov subspace represented by $\hat{\bf P}(i)$ is therefore similar to projecting $\hat{\bf a}_1(i)$. In our method, the estimation of $\hat{\bf d}(i)$ is separate from the update of $\hat{\bf a}_1(i)$, which means the steering vector estimation error used for the updates is obtained from $\hat{\bf d}(i)$, so that in the $k$th ($1{\leq}k<i$) snapshot, the error does not have to be orthogonal to $\hat{\bf a}_1(k)$, but should be orthogonal to another potentially better estimate $\hat{\bf a}_1(j)$ ($1{\leq}k<j{\leq}i$), resulting in a situation where the error is located inside the sector (see Fig. \ref{ipe3}). There are two benefits in the case which the error is inside the sector: faster convergence rate and smaller estimation error. We can obtain the infimum and supremum values in a similar way. By applying the inequality that the norm of the accumulated estimation error is no greater than the sum of the norms of all the piecewise estimation errors, we have
\begin{equation}
\lVert\hat{\bf e}(i)\rVert \leq \sum\limits_{k=1}^{i}\lVert{\bf a}_1\rVert\sin{\theta}_{i,k}, \label{eq103}
\end{equation}
where the parameters ${\theta}_{i,k}$ ($k=1,2,\dotsb,i$) satisfy the constraint in \eqref{eq94}. However, the right-hand side of \eqref{eq103} achieves its maximum value when all these parameters are equal (i.e., ${\theta}_{i,1}={\theta}_{i,2}=\dotsb={\theta}_{i,i}=\frac{{\theta}_i}{i}$) and it is given by
\begin{equation}
{\lVert\hat{\bf e}(i)\rVert}_{max}=i\lVert{\bf a}_1\rVert\sin\frac{{\theta}_i}{i}, \label{eq104}
\end{equation}
The right-hand side of \eqref{eq104} can be treated as a function of $i$ which is an increasing function on $i=1,2,\dotsb,\infty$. Therefore, we can take the limit of it to obtain the upper bound of ${\lVert\hat{\bf e}(i)\rVert}_{max}$, and so as to $\lVert\hat{\bf e}(i)\rVert$. In fact, when $i\rightarrow\infty$, the piecewise estimation error moves along the arc corresponding with ${\theta}_i$, resulting in the upper bound obtained is the same as the lower bound of the SQP method case, which is given by the right-hand side expression of \eqref{eq97} and defines the supremum of $\lVert\hat{\bf e}(i)\rVert$ in this case. Since we have already assumed that ${\theta}_i$ is less than $\pi/2$ so that $\hat{\bf e}(i)$ must be inside of the angular sector but its Euclidean norm cannot be smaller than the orthogonal distance between $\hat{\bf a}_1(0)$ to $\hat{\bf a}_1(i)$, so this orthogonal distance can define the lower bound of $\lVert\hat{\bf e}(i)\rVert$, which is actually the infimum and calculated by $\lVert{\bf a}_1\rVert\sin\theta_i$. Then, in the OKSPME method, ${\lVert\hat{\bf e}(i)\rVert}$ is bounded by
\begin{equation}
\inf\lVert\hat{\bf e}(i)\rVert = \lVert{\bf a}_1\rVert\sin\theta_i \leq {\lVert\hat{\bf e}(i)\rVert} < {\theta}_i\lVert{\bf a}_1\rVert = \sup\lVert\hat{\bf e}(i)\rVert. \label{eq105}
\end{equation}
By taking expectations of both \eqref{eq98} and \eqref{eq105}, we obtain
\begin{equation}
E[{\theta}_i]\lVert{\bf a}_1\rVert < \{E[{\lVert\hat{\bf e}(i)\rVert}]\}_{SQP} \leq \lVert{\bf a}_1\rVert\tan(E[{\theta}_i]), \label{eq106}
\end{equation}
\begin{equation}
\lVert{\bf a}_1\rVert\sin(E[\theta_{i}]) \leq \{E[{\lVert\hat{\bf e}(i)\rVert}]\}_{OKSPME} < E[{\theta}_i]\lVert{\bf a}_1\rVert. \label{eq107}
\end{equation}
On the other side, by substituting \eqref{eq98} and \eqref{eq105} in \eqref{eq85}, we obtain
\begin{equation}
0 \leq \{{\rm Var}[\lVert\hat{\bf e}(i)\rVert]\}_{SQP} \leq \frac{\lVert{\bf a}_1\rVert^2(\tan\theta_i-\theta_i)^2}{4}, \label{eq108}
\end{equation}
\begin{equation}
0 \leq \{{\rm Var}[\lVert\hat{\bf e}(i)\rVert]\}_{OKSPME} \leq \frac{\lVert{\bf a}_1\rVert^2(\theta_i-\sin\theta_i)^2}{4}. \label{eq109}
\end{equation}
Substituting \eqref{eq106}, \eqref{eq108} and \eqref{eq107}, \eqref{eq109} in \eqref{eq83}, respectively, we have
\begin{multline}
E^2[{\theta}_i]\lVert{\bf a}_1\rVert^2 < \{E[{\lVert\hat{\bf e}(i)\rVert}^2]\}_{SQP} \\
\leq \frac{\lVert{\bf a}_1\rVert^2(\tan\theta_i-\theta_i)^2}{4} + \lVert{\bf a}_1\rVert^2\tan^2(E[{\theta}_i]), \label{eq110}
\end{multline}
\begin{multline}
\lVert{\bf a}_1\rVert^2\sin^2(E[\theta_{i}]) \leq \{E[{\lVert\hat{\bf e}(i)\rVert}^2]\}_{OKSPME} \\
< \frac{\lVert{\bf a}_1\rVert^2(\theta_i-\sin\theta_i)^2}{4} + E^2[{\theta}_i]\lVert{\bf a}_1\rVert^2. \label{eq111}
\end{multline}
However, $E[{\theta}_i]$ also has its lower and upper bounds. Since our analysis focuses on the unbiased case only as mentioned, the true steering vector is located in the center of the angular sector and the estimate $\hat{\bf a}_1(i)$ is always closer to the center than $\hat{\bf a}_1(0)$. Let us assume that even if the estimate $\hat{\bf a}_1(i)$ always happens to be very close to either edge of the sector, no matter how $\hat{\bf a}_1(0)$ is chosen within the sector, ${\theta}_i$ will vary from $0$ to $2\theta$ with equal probability, or equivalently, uniformly distributed within $[0,2\theta)$, in which case we can obtain the upper bound for $E[{\theta}_i]$ by taking the average between $0$ to $2\theta$, which is obtained as $\theta$. On the other hand, if we assume that the estimate $\hat{\bf a}_1(i)$ always happens to be exactly at the center of the sector, resulting in that ${\theta}_i$ can only vary from $0$ to $\theta$, or uniformly distributed within $[0,\theta]$ in which case $E[{\theta}_i]=\theta/2$, resulting in the lower bound of $E[{\theta}_i]$ is $\theta/2$.  Therefore, the upper and lower bounds for ${\rm MSE}\{\hat{\bf a}_1(i)\}$ can be further obtained by substituting $E[{\theta}_i]_{max}\rightarrow\theta$, $[\theta_i]_{max}\rightarrow2\theta$ and $E[{\theta}_i]_{min}=\theta/2$, $[\theta_i]_{min}=0$ into the upper and lower bounds of \eqref{eq110} and \eqref{eq111} respectively, resulting in
\begin{multline}
\frac{\theta^2}{4}\lVert{\bf a}_1\rVert^2 < \{E[{\lVert\hat{\bf e}(i)\rVert}^2]\}_{SQP} \\
< \frac{\lVert{\bf a}_1\rVert^2(\tan2\theta-2\theta)^2}{4} + \lVert{\bf a}_1\rVert^2\tan^2\theta, \label{eq112}
\end{multline}
\begin{multline}
\lVert{\bf a}_1\rVert^2\sin^2\frac{\theta}{2} \leq \{E[{\lVert\hat{\bf e}(i)\rVert}^2]\}_{OKSPME} \\
< \frac{\lVert{\bf a}_1\rVert^2(2\theta-\sin2\theta)^2}{4} + \theta^2\lVert{\bf a}_1\rVert^2. \label{eq113}
\end{multline}
Finally, by combining the expectation of the mean-squared initial guess error $E[\lVert{\bf\epsilon}\rVert^2]$ in \eqref{eq93} with \eqref{eq112} and \eqref{eq113}, we obtain the bounds for the MSE of the steering vector estimate ${\rm MSE}\{\hat{\bf a}_1(i)\}$ as
\begin{multline}
(2-\frac{2\sin\theta}{\theta}+\frac{\theta^2}{4})\lVert{\bf a}_1\rVert^2 < \{{\rm MSE}\{\hat{\bf a}_1(i)\}\}_{SQP} \\
< (2-\frac{2\sin\theta}{\theta}+\frac{(\tan2\theta-2\theta)^2}{4}+\tan^2\theta)\lVert{\bf a}_1\rVert^2, \label{eq114}
\end{multline}
\begin{multline}
(2-\frac{2\sin\theta}{\theta}+\sin^2\frac{\theta}{2})\lVert{\bf a}_1\rVert^2 \leq \{{\rm MSE}\{\hat{\bf a}_1(i)\}\}_{OKSPME} \\
< (2-\frac{2\sin\theta}{\theta}+\frac{(2\theta-\sin2\theta)^2}{4}+\theta^2)\lVert{\bf a}_1\rVert^2. \label{eq115}
\end{multline}
From \eqref{eq114} and \eqref{eq115}, we can see that the MSEs now only depend on two parameters: the norm of the true steering vector and the angular sector spread. The lower and upper bounds of the proposed OKSPME method are lower than those of the SQP method. As mentioned before, it is important that the presumed angular sector spread $2\theta$ must be less than $\pi/2$ (i.e., $90^{\circ}$) to ensure the previous assumption of ${\theta}_i<\pi/2$ is always valid.

\subsection{Complexity Analysis}

\begin{figure}[!htb]
\begin{center}
\def\epsfsize#1#2{0.95\columnwidth}
\epsfbox{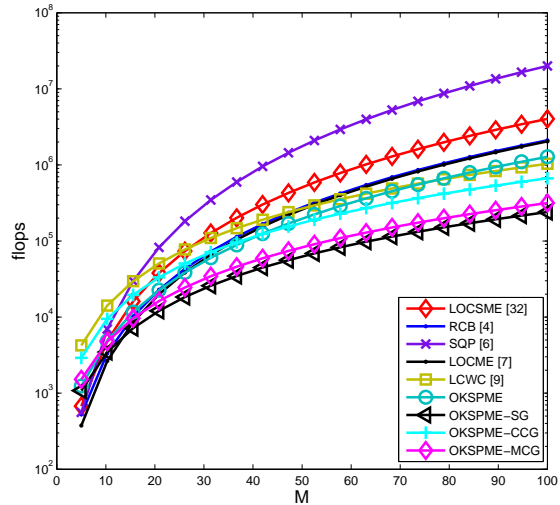}
\caption{Complexity Comparison} \label{figure1}
\end{center}
\end{figure}

The computational complexity analysis is discussed in this subsection. We measure the total number of additions and multiplications (i.e., flops) in terms of the number of sensors $M$ performed for each snapshot for the proposed algorithms and the existing ones and list them in Table \ref{table5}. Note that the SQP method in \cite{r6} has a highly-variant computational complexity in different snapshots, due to the online optimization program based on random choices of the presumed steering vector. However, it is usually in ${\mathcal O}(M^{3.5})$. The complexity of the LCWC algorithm in \cite{r9} often requires a much larger $n$ than that in the proposed LOCSME-CCG algorithm. It is obvious that all of the proposed algorithms have their complexity depending on the Krylov subspace model order $m$, which is determined from Table \ref{table1} and is no larger than $K+1$. For the convenience of comparison, we eliminate all parameters except $M$ by setting them to common values (the values of $n$ in LCWC and OKSPME-CCG is set to $50$ and $5$ respectively, $m=K+1$ where $K=3$) and illustrate their complexity with $M$ varying from $10$ to $100$ as shown in Fig. \ref{figure1}. As can be seen that the proposed OKSPME-SG and OKSPME-MCG algorithms have lower complexity than the other algorithms.

\begin{table}
\small
\begin{center}
\caption{Complexity Comparison}
\begin{tabular}{|c|c|}
\hline
RAB Algorithms & Flops \\
\hline
LOCSME \cite{r32} & $4M^3+3M^2+20M$ \\
\hline
RCB \cite{r4} & $2M^3+11M^2$ \\
\hline
SQP \cite{r6} & ${\mathcal O}(M^{3.5})$ \\
\hline
LOCME \cite{r7} & $2M^3+4M^2+5M$ \\
\hline
LCWC \cite{r9} & $2nM^2+7nM$ \\
\hline
OKSPME & \begin{tabular}[c]{@{}c@{}} $M^3+(4m+11)M^2$ \\ $+(3m^2+5m+20)M$ \end{tabular}  \\
\hline
OKSPME-SG & \begin{tabular}[c]{@{}c@{}} $(4m+7)M^2$ \\ $+(3m^2+5m+33)M$ \end{tabular} \\
\hline
OKSPME-CCG & \begin{tabular}[c]{@{}c@{}} $(4m+8n+8)M^2$ \\ $+(3m^2+5m+33n+29)M$ \end{tabular} \\
\hline
OKSPME-MCG & \begin{tabular}[c]{@{}c@{}} $(4m+14)M^2$ \\ $+(3m^2+5m+86)M$ \end{tabular} \\
\hline
\end{tabular} \label{table5}
\end{center}
\end{table}

\section{Simulations}

In this section, we present and discuss the simulation results of the proposed RAB algorithms by comparing them to some of the existing RAB techniques. We consider uniform linear arrays (ULA) of omnidirectional sensors with half wavelength spacing. To produce all the figures (if unspecified in a few scenario), $100$ repetitions are executed to obtain each point of the curves and a maximum of $i=300$ snapshots are observed. The desired signal is assumed to arrive at ${\theta}_1=10^\circ$. The signal-to-interference ratio (SIR) is fixed at $0$dB. As the prior knowledge, the angular sector in which the desired signal is assumed to be located is chosen as $[{\theta}_1-5^\circ,{\theta}_1+5^\circ]$. The results focus on the beamformer output SINR performance versus the number of snapshots, or a variation of input SNR ($-10$dB to $30$dB) and both coherent and incoherent local scattering mismatch \cite{r5} scenarios are considered.

\subsection{{Mismatch due to Coherent Local Scattering}}

All simulations in this subsection consider coherent local scattering. With time-invariant coherent local scattering, the steering vector of the desired signal is modeled as
\begin{equation}
{\bf a}_1={\bf p}+\sum\limits_{k=1}^4{e^{j{\varphi}_k}}{\bf b}({\theta}_k), \label{eq116}
\end{equation}
where ${\bf p}$ corresponds to the direct path while ${\bf b}({\theta}_k)(k=1,2,3,4)$ corresponds to the scattered paths. The angles ${\theta}_k(k=1, 2, 3, 4)$ are randomly and independently drawn in each simulation run from a uniform generator with mean $10^\circ$ and standard deviation $2^\circ$. The angles ${\varphi}_k(k=1, 2, 3, 4)$ are independently and uniformly taken from the interval $[0,2\pi]$ in each simulation run. Both ${\theta}_k$ and ${\varphi}_k$ change from trials while remaining constant over snapshots.

\begin{figure}[!htb]
\begin{center}
\def\epsfsize#1#2{0.95\columnwidth}
\epsfbox{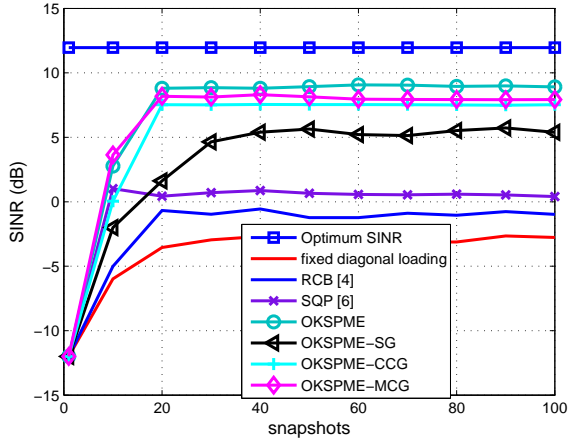}
\caption{Coherent local scattering, SINR versus snapshots, $M=10$, $K=3$, ${\rm INR}=20{\rm dB}$} \label{figure-add}
\end{center}
\end{figure}

We firstly compare our proposed methods with some classical RAB methods (i.e., standard diagonal loading method with a fixed loading factor equal to $10$ times the noise variance, the RCB method in \cite{r4} which estimates the loading factor iteratively, and the method that solves an online quadratic optimization programming, which refers to the SQP method \cite{r6}). The numbers of sensors and signal sources (including the desired signal) are set to $M=10$ and $K=3$, respectively. For this case only, we set the INR (interferences-to-noise ratio) to $20$dB and illustrate the SINR performance versus snapshots within $100$ snapshots in Fig. \ref{figure-add}. The two interferers are arranged to be in the directions of ${\theta}_2=30^\circ$ and ${\theta}_3=50^\circ$, respectively. The other user-defined parameters, if unspecified, (e.g. the step size $\mu$ and the forgetting factor $\lambda$) are manually optimized to give the best algorithm performance, which is also applied for the other simulation scenarios.

\begin{figure}[!htb]
\begin{center}
\def\epsfsize#1#2{0.95\columnwidth}
\epsfbox{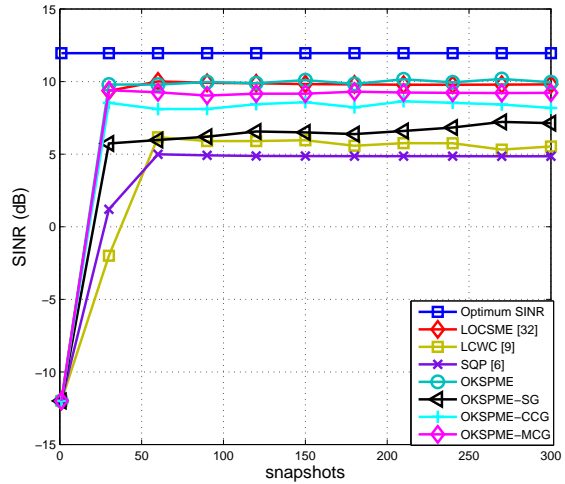}
\caption{Coherent local scattering, SINR versus snapshots, $M=12$, $K=3$} \label{figure2}
\end{center}
\end{figure}

\begin{figure}[!htb]
\begin{center}
\def\epsfsize#1#2{0.95\columnwidth}
\epsfbox{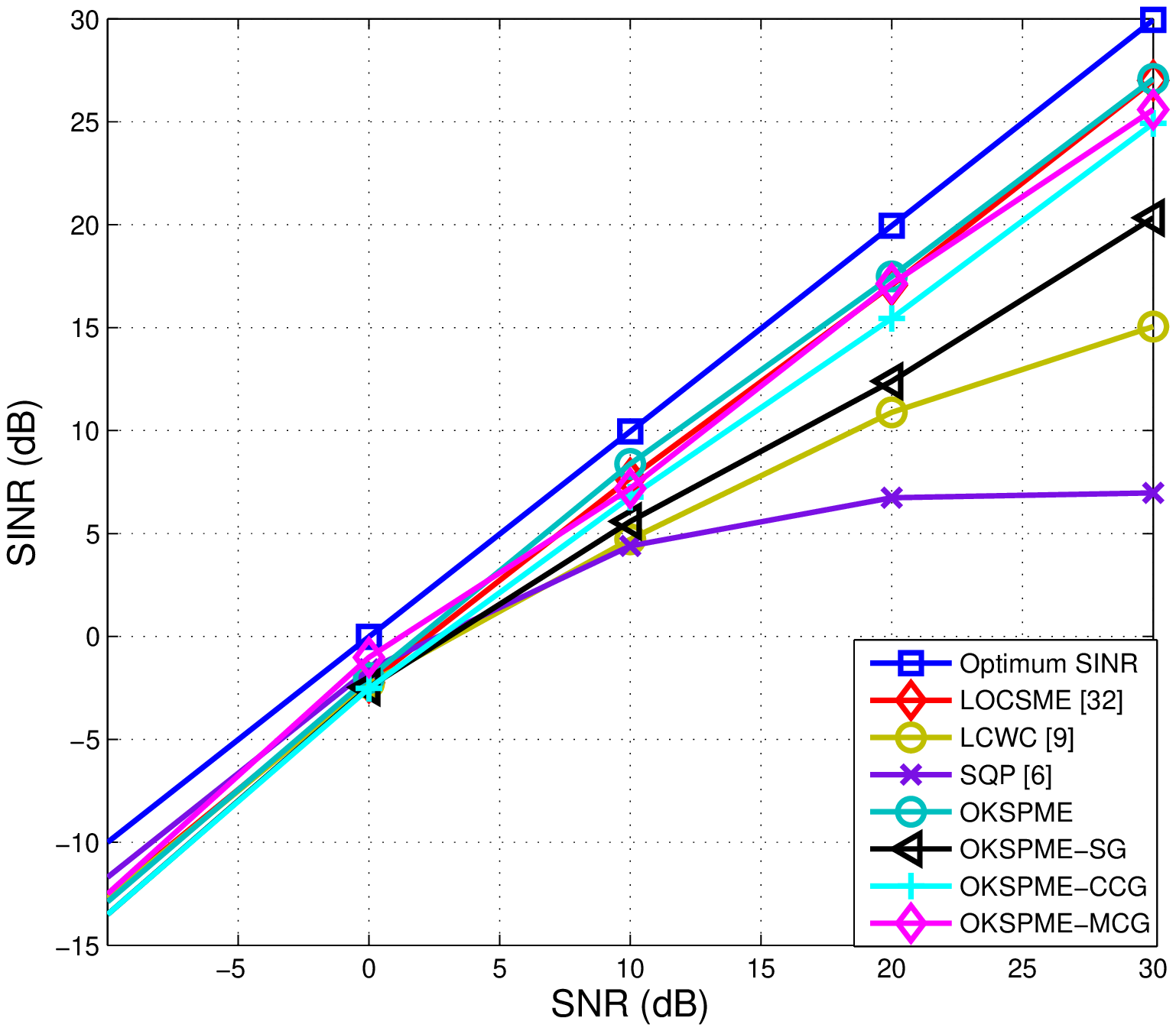}
\caption{Coherent local scattering, SINR versus SNR, $M=12$, $K=3$} \label{figure3}
\end{center}
\end{figure}

We then set the number of sensors to $M=12$, the number of signal sources as (including the desired signal) $K=3$ and illustrate the SINR versus snapshots and the SINR versus input SNR performance in Fig. \ref{figure2} and Fig. \ref{figure3} respectively. The two interferers are arranged to be in the directions of ${\theta}_2=30^\circ$ and ${\theta}_3=50^\circ$, respectively. In either Fig. \ref{figure2} or Fig. \ref{figure3}, we can see that the proposed OKSPME method has a very similar or slightly better performance compared to the LOCSME algorithm of \cite{r32} and both of them have the best performance. Furthermore, the proposed OKSPME-CCG and OKSPME-MCG algorithms also achieve very close performance to OKSPME.

\begin{figure}[!htb]
\begin{center}
\def\epsfsize#1#2{0.95\columnwidth}
\epsfbox{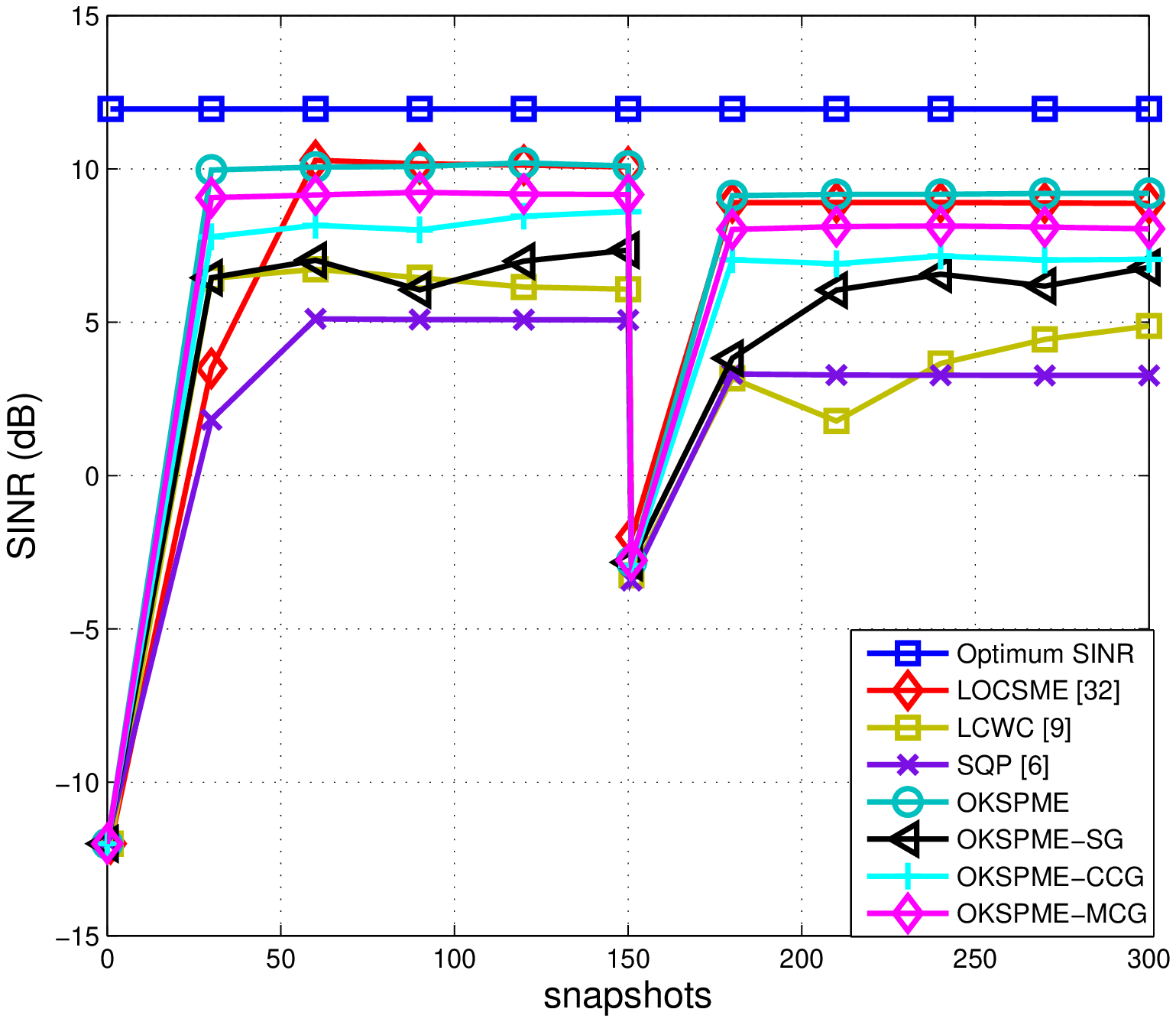}
\caption{Coherent local scattering, SINR versus snapshots, $M=12$} \label{figure4}
\end{center}
\end{figure}

In Fig. \ref{figure4}, we assess the SINR performance versus snapshots of those selected algorithms in a specific time-varying scenario which encounters a halfway redistribution of the interferers at a certain snapshot. In this case, the number of sensors is kept at $M=12$, whereas the details of the interferers are given in Table \ref{table6}.  

\begin{table}
\small
\begin{center}
\caption{Changes of Interferers}
\begin{tabular}{|c|c|c|}
\hline
Snapshots & \begin{tabular}[c]{@{}c@{}} Number of Interferers \\ ($K-1$) \end{tabular} & DoAs\\
\hline
$0-150$ & $2$ & ${\theta}_2=30^\circ$, ${\theta}_3=50^\circ$. \\
\hline
$151-300$ & $5$ & \begin{tabular}[c]{@{}c@{}} ${\theta}_2=20^\circ$, ${\theta}_3=30^\circ$, ${\theta}_4=40^\circ$, \\ ${\theta}_5=50^\circ$, ${\theta}_6=60^\circ$. \end{tabular} \\
\hline
\end{tabular} \label{table6}
\end{center}
\end{table}

\begin{figure}[!htb]
\begin{center}
\def\epsfsize#1#2{0.95\columnwidth}
\epsfbox{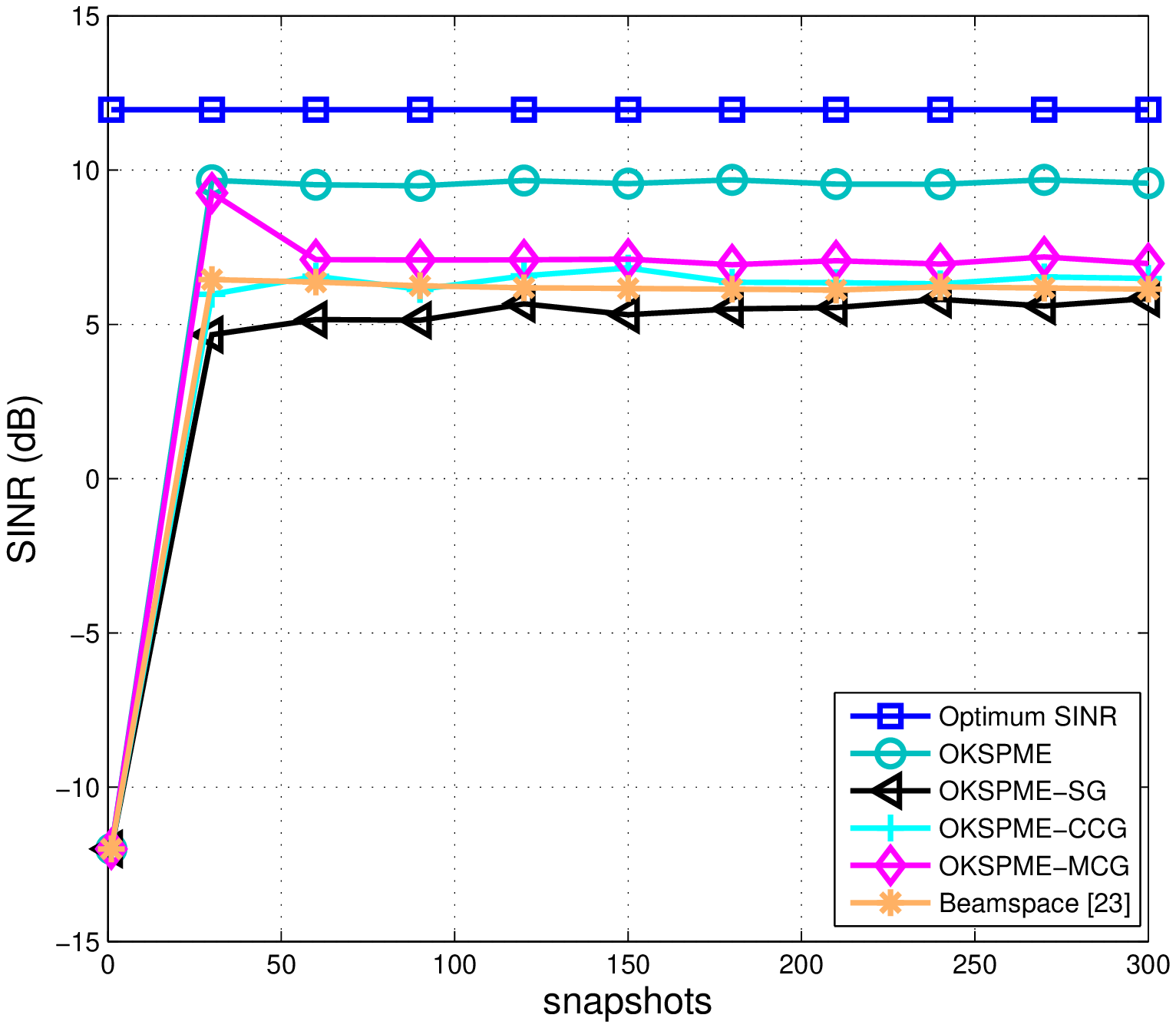}
\caption{Coherent local scattering, SINR versus snapshots, $M=40$, $K=3$} \label{figure5}
\end{center}
\end{figure}

\begin{figure}[!htb]
\begin{center}
\def\epsfsize#1#2{0.95\columnwidth}
\epsfbox{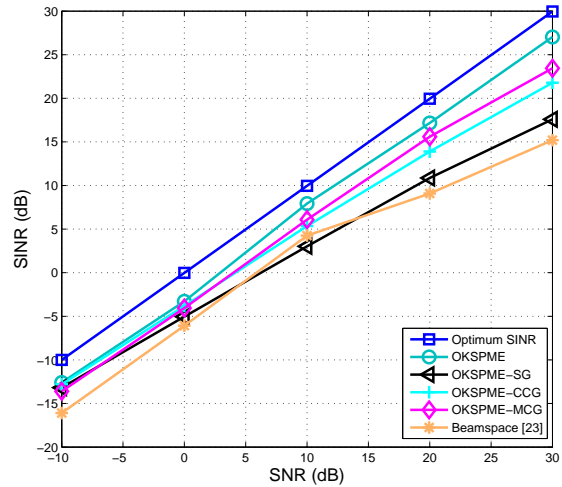}
\caption{Coherent local scattering, SINR versus SNR, $M=40$, $K=3$} \label{figure6}
\end{center}
\end{figure}

In Figs. \ref{figure5} and \ref{figure6}, we set the number of signal sources to $K=3$, but increase the number of sensors from $M=12$ to $M=40$ and study the SINR versus snapshots and the SINR versus input SNR performance of the selected and proposed dimensionality reduction RAB algorithms, respectively. We set the reduced-dimension as $D=4$ for the beamspace based algorithm \cite{r23} in all simulations. This time, it is clear that the proposed OKSPME, OKSPME-SG, OKSPME-CCG and OKSPME-MCG algorithms all have a certain level of performance degradation compared to the scenario where $M=12$. The proposed OKSPME based algorithms achieve better performances than the beamspace approach.

\subsection{{Mismatch due to Incoherent Local Scattering}}

\begin{figure}[!htb]
\begin{center}
\def\epsfsize#1#2{0.95\columnwidth}
\epsfbox{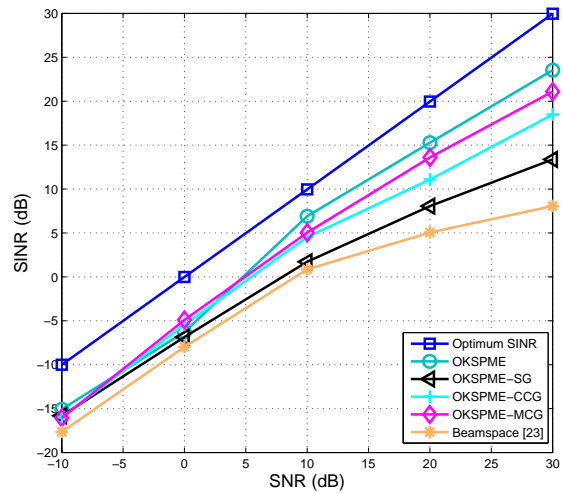}
\caption{Incoherent local scattering, SINR versus SNR, $M=40$, $K=3$} \label{figure7}
\end{center}
\end{figure}

In this case, the desired signal affected by incoherent local scattering has a time-varying signature and its steering vector is modeled by
\begin{equation}
{\bf a}_1(i)=s_0(i){\bf p}+\sum\limits_{k=1}^4{s_k(i)}{\bf b}({\theta}_k), \label{eq117}
\end{equation}
where $s_k(i)(k=0, 1, 2, 3, 4)$ are i.i.d zero mean complex Gaussian random variables independently drawn from a random generator. The angles ${\theta}_k(k=0, 1, 2, 3, 4)$ are drawn independently in each simulation run from a uniform generator with mean $10^\circ$ and standard deviation $2^\circ$. At this time, $s_k(i)$ changes both from run to run and from snapshot to snapshot. In order to show the effects caused by incoherent scattering only, we set the parameters $M=40$ and $K=3$, study the SINR versus SNR performance of the selected algorithms in Fig. \ref{figure7} and compare the results with Fig. \ref{figure6}. As a result, a performance degradation is observed for all the studied algorithms. This is because the time-varying nature of incoherent scattering results in more dynamic and environmental uncertainties in the system, which increases the steering vector mismatch.

\section{Conclusion}

We have proposed the OKSPME algorithm based on the exploitation of cross-correlation mismatch estimation and the orthogonal Krylov subspace. In addition, low complexity RAB algorithms, OKSPME-SG, OKSPME-CCG and OKSPME-MCG have been developed to enable the beamforming weights to be updated recursively without matrix inversions. A detailed steering vector estimation MSE analysis for the general RAB design approach that relies on a presumed angular sector as prior knowledge has been provided. The computational complexity of the proposed and some of the existing algorithms have been compared and discussed. Simulation results have shown that the proposed algorithms have robustness against different choices of user-defined parameters and environmental effects, and achieved excellent output SINR performance especially in medium-high input SNR values.


\begin{thebibliography}{10}
{\footnotesize
{\linespread{1.0}

\bibitem{r1}
H. L. Van Trees, {\it Optimum Array Processing}, New York: Wiley, 2002.

\bibitem{r2}
S. A. Vorobyov, A. B. Gershman and Z. Luo, ``Robust Adaptive Beamforming using Worst-Case Performance Optimization: A Solution to the Signal Mismatch Problem," \emph{IEEE Transactions on Signal Processing}, Vol. 51, No. 4, pp 313-324, Feb 2003.

\bibitem{r3}
A. Khabbazibasmenj, S. A. Vorobyov and A. Hassanien, ``Robust Adaptive Beamforming Based on Steering Vector Estimation with as Little as Possible Prior Information," \emph{IEEE Transactions on Signal Processing}, Vol. 60, No. 6, pp 2974-2987, June 2012.

\bibitem{r4}
J. Li, P. Stoica and Z. Wang, ``On Robust Capon Beamforming and Diagonal Loading," \emph{IEEE Transactions on Signal Processing}, Vol. 57, No. 7, pp 1702-1715, July 2003.

\bibitem{r5}
D. Astely and B. Ottersten, ``The effects of Local Scattering on Direction of Arrival Estimation with Music," \emph{IEEE Transactions on Signal Processing}, Vol. 47, No. 12, pp 3220-3234, Dec 1999.

\bibitem{r6}
A. Hassanien, S. A. Vorobyov and K. M. Wong, ``Robust Adaptive Beamforming Using Sequential Quadratic Programming: An Iterative Solution to the Mismatch Problem," \emph{IEEE Sig. Proc. Letters.}, Vol. 15, pp 733-736, 2008.

\bibitem{r7}
L. Landau, R. de Lamare, M. Haardt, ``Robust Adaptive Beamforming Algorithms Using Low-Complexity Mismatch Estimation," \emph{Proc. IEEE Statistical Signal Processing Workshop}, 2011.

\bibitem{r8}
J. Zhuang and A. Manikas, ``Interference cancellation beamforming robust to pointing errors," IET Signal Process., Vol. 7, No. 2, pp. 120-127, April 2013.

\bibitem{r9}
A. Elnashar, ``Efficient implementation of robust adaptive beamforming based on worst-case performance optimization," IET Signal Process., Vol. 2, No. 4, pp. 381-393, Dec 2008.

\bibitem{r10}
L. Wang and R. C. de Lamare, ``Constrained adaptive filtering algorithms based on conjugate gradient techniques for beamforming," IET Signal Process., Vol. 4, No. 6, pp. 686-697, Feb 2010.

\bibitem{r11}
J. P. Lie, W. Ser and C. M. S. See, ``Adaptive Uncertainty Based Iterative Robust Capon Beamformer Using Steering Vector Mismatch Estimation," \emph{IEEE Transactions on Signal Processing}, Vol. 59, No. 9, Sep 2011.

\bibitem{r12}
A. Filimon, ``Krylov Subspace Iteration Methods," May 2008.

\bibitem{r13}
H. A. van der Vorst, ``Iterative Krylov Methods for Large Linear Systems," Cambridge University Press, 2003.

\bibitem{r14}
M. Morelli, L. Sangunietti and U. Mengali, ``Channel Estimation for Adaptive Frequency-Domain Equalization," \emph{IEEE Transactions on Wireless Communications}, Vol. 4, No. 5, pp. 53-57, Sep 2005.

\bibitem{r15}
Z. Bai, ``Krylov subspace techniques for reduced-order modeling of large-scale dynamical systems," Applied Numerical Mathematics, Department of Computer Science, University of California, 2002.

\bibitem{r16}
R. C. de Lamare, M. Haardt and R. Sampaio-Neto, ``Blind Adaptive Constrained Reduced-Rank Parameter Estimation Based on Constant Modulus Design for CDMA Interference Suppression," \emph{IEEE Transactions on Signal Processing}, Vol. 56, No. 6, June 2008.

\bibitem{r17}
H. Ge, I. P. Kirsteins and L. L. Scharf, ``Data Dimension Reduction Using Krylov Subspaces Making Adaptive Beamformers Robust to Model Order-Determination," in \emph{Int. Conf. Acoustics, Speech, and Signal Processing (ICASSP)}, Vol. 4, 2006.

\bibitem{r18}
I. P. Kirsteins and H. Ge, ``Performance Analysis of Krylov Space Adaptive Beamformers," \emph{Sensor Array and Multichannel Processing, Fourth IEEE Workshop}, pp. 16-20, July 2006.

\bibitem{r19}
S. D. Somasundaram, N. H. Parsons, P. Li and R. C. de Lamare, ``Data-Adaptive Reduced-Dimension Robust Capon Beamforming," \emph{Int. Conf. Acoustics, Speech, and Signal Processing (ICASSP)}, pp. 4159-4163, May 2013.

\bibitem{r20}
S. D. Somasundaram, N. H. Parsons, P. Li and R. C. de Lamare, ``Reduced-Dimension Robust Capon Beamforming Using Krylov-Subspace Techniques," \emph{IEEE Transactions on Aerospace and Electronic Systems}, Vol. 51, No. 1, pp. 270-289, Jan 2015.

\bibitem{r21}
V. Druskin and V. Simoncini, ``Adaptive Rational Krylov Subspaces for Large-Scale Dylymical Systems," \emph{Systems and Control Letters}, Vol. 60, No. 8, pp. 546-560, 2011.

\bibitem{r22}
K. Carlberg and C. Farhat, ``An Adaptive POD-Krylov Reduced-Order Model for Structural Optimization," \emph{8th World Congress on Structural and Multidisciplinary Optimization}, June 2009.

\bibitem{r23}
A. Hassanien and S. A. Vorobyov, ``A Robust Adaptive Dimension Reduction Technique With Application to Array Processing," \emph{IEEE Sig. Proc. Letters.} Vol. 16, No. 1, pp. 22-25, Jan 2009.

\bibitem{r24}
M. Yukawa, R. C. de Lamare and I. Yamada, ``Robust Reduced-Rank Adaptive Algorithm Based on Parallel Subgradient Projection and Krylov Subspace," \emph{IEEE Transactions on Signal Processing}, Vol. 57, No. 12, Dec 2009.

\bibitem{r25}
C. Dumard and T. Zemen, ``Double Krylov Subspace Approximation for Low Complexity Iterative Multi-User Decoding and Time-Variant Channel Estimation," \emph{6th IEEE Workshop on Signal Processing Advances in Wireless Communications}, pp. 328-332, June 2005.

\bibitem{r26}
R. C. de Lamare and R. Sampaio-Neto, ``Reduced-Rank Adaptive Filtering Based on Joint Iterative Optimization of Adaptive Filters," \emph{IEEE Sig. Proc. Letters.} Vol. 14, No. 12, Dec 2007.

\bibitem{r27}
N. Song, W. U. Alokozai, R. C. de Lamare and M. Haardt, ``Adaptive Widely Linear Reduced-Rank Beamforming Based on Joint Iterative Optimization," \emph{IEEE Sig. Proc. Letters.} Vol. 21, No. 3, Mar 2014.

\bibitem{r28}
S. D. Somasundaram, ``A Framework for Reduced Dimension Robust Capon Beamforming," \emph{IEEE Statistical Signal Processing Workshop (SSP)}, 2011.

\bibitem{r29}
P. Li and R. C. de Lamare, ``Low-Complexity Robust Data-Dependant Dimensionality Reduction Based on Joint Iterative Optimization of Parameters," \emph{IEEE 5th International Workshop on Computational Advances in Multi-Sensor Adaptive Processing (CAMSAP)}, pp. 49-52, Dec 2013.

\bibitem{r30}
R. C. de Lamare and R. Sampaio-Neto, ``Adaptive Reduced-Rank Processing Based on Joint and Iterative Interpolation, Decimation, and Filtering," \emph{IEEE Transactions on Signal Processing}, Vol. 57, No. 7, July 2009.

\bibitem{r31}
L. Wang, R. C. de Lamare and M. Haardt, ``Direction finding Algorithms with Joint Iterative Subspace Optimization," \emph{IEEE Transactions on Aerospace and Electronic Systems}, Vol. 50, No. 4, pp. 2541-2553, Oct 2014.

\bibitem{r32}
H. Ruan and R. C. de Lamare, ``Robust Adaptive Beamforming Using a Low-Complexity Shrinkage-Based Mismatch Estimation Algorithm," \emph{IEEE Sig. Proc. Letters.}, Vol. 21, No. 1, pp. 60-64, Nov 2013.

\bibitem{r33}
C. Zhang, R. Tian, X. Tan, ``A Leakage-Based Dimensionality Reduction Beamforming for MIMO Cognitive Radio Networks," \emph{8th International Conference on Communications and Networking in China (CHINACOM)}, 2013.

\bibitem{r34}
R. Fa, R. C. de Lamare and D. Zanatta-Filho, ``Reduced-Rank STAP Algorithm for Adaptive Radar Based on Joint Iterative Optimization of Adaptive Filters," \emph{42nd Asilomar Conference on Signals, Systems and Computers}, pp. 533-537, Oct 2008.

\bibitem{r35}
R. Fa, R. C. de Lamare and P. Clarke, ``Reduced-Rank STAP for MIMO Radar Based on Joint Iterative Optimization of Knowledge-Aided Adaptive Filters," \emph{Conference Record of the Forty-Third Asilomar Conference on Signals, Systems and Computers}, pp. 496-500, Nov 2009.

\bibitem{r36}
S.D. Somasundaram, ``Reduced dimension robust Capon beamforming for large aperture passive sonar arrays," IET Radar, Sonar and Navigation, Vol. 5, No. 7, pp. 707-715, 2011.

\bibitem{r37}
Popoviciu, ``Sur Les Equations Algebriques Ayant Toutes Leurs Racines Réelles," Mathematica 9, pp. 129-145, 1935.

\bibitem{scharf}
L. L. Scharf and D. W. Tufts, ``Rank reduction for modeling
stationary signals," \textit{IEEE Transactions on Acoustics, Speech
and Signal Processing}, vol. ASSP-35, pp. 350-355, March 1987.



\bibitem{bar-ness} A. M. Haimovich
and Y. Bar-Ness, ``An eigenanalysis interference canceler," {\it
IEEE Trans. on Signal Processing}, vol. 39, pp. 76-84, Jan. 1991.

\bibitem{pados99} D. A. Pados and S. N. Batalama "Joint space-time
auxiliary vector filtering for DS/CDMA systems with antenna arrays"
\textit{ IEEE Transactions on Communications}, vol. 47, no. 9, pp.
1406 - 1415, 1999.



\bibitem{reed98} J. S. Goldstein, I. S. Reed and L. L. Scharf
"A multistage representation of the Wiener filter based on
orthogonal projections" \textit{IEEE Transactions on Information
Theory}, vol. 44, no. 7, 1998.

\bibitem{hua}
Y. Hua, M. Nikpour and P. Stoica, "Optimal reduced rank estimation
and filtering," IEEE Transactions on Signal Processing, pp. 457-469,
Vol. 49, No. 3, March 2001.


\bibitem{goldstein}
M. L. Honig and J. S. Goldstein, ``Adaptive reduced-rank
interference suppression based on the multistage Wiener filter,"
\textit{ IEEE Transactions on Communications}, vol. 50, no. 6, June
2002.

\bibitem{santos}
E. L. Santos and M. D. Zoltowski, ``On Low Rank MVDR Beamforming
using the Conjugate Gradient Algorithm", \textit{Proc. IEEE
International Conference on Acoustics, Speech and Signal
Processing}, 2004.

\bibitem{qian}
Q. Haoli and S.N. Batalama, ``Data record-based criteria for the
selection of an auxiliary vector estimator of the MMSE/MVDR filter",
\textit{IEEE Transactions on Communications}, vol. 51, no. 10, Oct.
2003, pp. 1700 - 1708.

\bibitem{delamarespl07}
R. C. de Lamare and R. Sampaio-Neto, ``Reduced-Rank Adaptive
Filtering Based on Joint Iterative Optimization of Adaptive
Filters", \textit{IEEE Signal Processing Letters}, Vol. 14, no. 12,
December 2007.

\bibitem{xutsa}
Z. Xu and M.K. Tsatsanis, ``Blind adaptive algorithms for minimum
variance CDMA receivers," \textit{IEEE Trans. Communications}, vol.
49, No. 1, January 2001.

\bibitem{delamaretsp}
R. C. de Lamare and R. Sampaio-Neto, ``Low-Complexity Variable
Step-Size Mechanisms for Stochastic Gradient Algorithms in Minimum
Variance CDMA Receivers", \textit{IEEE Trans. Signal Processing},
vol. 54, pp. 2302 - 2317, June 2006.

\bibitem{kwak}
C. Xu, G. Feng and K. S. Kwak, ``A Modified Constrained Constant
Modulus Approach to Blind Adaptive Multiuser Detection," \textit{
IEEE Trans. Communications}, vol. 49, No. 9, 2001.

\bibitem{xu&liu}
Z. Xu and P. Liu, ``Code-Constrained Blind Detection of CDMA Signals
in Multipath Channels," \textit{ IEEE Sig. Proc. Letters}, vol. 9,
No. 12, December 2002.

%
\bibitem{delamareccm}
R. C. de Lamare and R. Sampaio Neto, "Blind Adaptive
Code-Constrained Constant Modulus Algorithms for CDMA Interference
Suppression in Multipath Channels", \textit{ IEEE Communications
Letters}, vol 9. no. 4, April, 2005.

\bibitem{wcccm}
L. Landau, R. C. de Lamare and M. Haardt, ``Robust adaptive
beamforming algorithms using the constrained constant modulus
criterion," IET Signal Processing, vol.8, no.5, pp.447-457, July
2014.

\bibitem{delamareelb}
R. C. de Lamare, ``Adaptive Reduced-Rank LCMV Beamforming Algorithms
Based on Joint Iterative Optimisation of Filters",
\textit{Electronics Letters}, vol. 44, no. 9, 2008.


\bibitem{jidf}
R. C. de Lamare and R. Sampaio-Neto, ``Adaptive Reduced-Rank
Processing Based on Joint and Iterative Interpolation, Decimation
and Filtering", \textit{IEEE Transactions on Signal Processing},
vol. 57, no. 7, July 2009, pp. 2503 - 2514.

\bibitem{delamarecl}
R. C. de Lamare and Raimundo Sampaio-Neto, ``Reduced-rank
Interference Suppression for DS-CDMA based on Interpolated FIR
Filters", \textit{IEEE Communications Letters}, vol. 9, no. 3, March
2005.

\bibitem{delamaresp}
R. C. de Lamare and R. Sampaio-Neto, ``Adaptive Reduced-Rank MMSE
Filtering with Interpolated FIR Filters and Adaptive Interpolators",
\textit{IEEE Signal Processing Letters}, vol. 12, no. 3, March,
2005.

\bibitem{delamaretvt}
R. C. de Lamare and R. Sampaio-Neto, ``Adaptive Interference
Suppression for DS-CDMA Systems based on Interpolated FIR Filters
with Adaptive Interpolators in Multipath Channels", \textit{IEEE
Trans. Vehicular Technology}, Vol. 56, no. 6, September 2007.

\bibitem{jioel}
R. C. de Lamare, ``Adaptive Reduced-Rank LCMV Beamforming Algorithms
Based on Joint Iterative Optimisation of Filters," Electronics
Letters, 2008.


\bibitem{delamarespl07}
R. C. de Lamare and R. Sampaio-Neto, ``Reduced-rank adaptive
filtering based on joint iterative optimization of adaptive
filters",  \textit{IEEE Signal Process. Lett.}, vol. 14, no. 12, pp.
980-983, Dec. 2007.

\bibitem{delamare_ccmmswf}
R. C. de Lamare, M. Haardt, and R. Sampaio-Neto, ``Blind Adaptive
Constrained Reduced-Rank Parameter Estimation based on Constant
Modulus Design for CDMA Interference Suppression", \textit{IEEE
Transactions on Signal Processing}, June 2008.

\bibitem{jidf_echo}
M. Yukawa, R. C. de Lamare and R. Sampaio-Neto, ``Efficient Acoustic
Echo Cancellation With Reduced-Rank Adaptive Filtering Based on
Selective Decimation and Adaptive Interpolation," IEEE Transactions
on Audio, Speech, and Language Processing, vol.16, no. 4, pp.
696-710, May 2008.

\bibitem{delamaretvt10}
R. C. de Lamare and R. Sampaio-Neto, ``Reduced-rank space-time
adaptive interference suppression with joint iterative least squares
algorithms for spread-spectrum systems," \textit{IEEE Trans. Vehi.
Technol.}, vol. 59, no. 3, pp. 1217-1228, Mar. 2010.

\bibitem{delamaretvt2011ST}
R. C. de Lamare and R. Sampaio-Neto, ``Adaptive reduced-rank
equalization algorithms based on alternating optimization design
techniques for MIMO systems," \textit{IEEE Trans. Vehi. Technol.},
vol. 60, no. 6, pp. 2482-2494, Jul. 2011.


\bibitem{delamare10}
R. C. de Lamare, L. Wang, and R. Fa, ``Adaptive reduced-rank LCMV
beamforming algorithms based on joint iterative optimization of
filters: Design and analysis," Signal Processing, vol. 90, no. 2,
pp. 640-652, Feb. 2010.

\bibitem{fa10}
R. Fa, R. C. de Lamare, and L. Wang, ``Reduced-Rank STAP Schemes for
Airborne Radar Based on Switched Joint Interpolation, Decimation and
Filtering Algorithm," \textit{IEEE Transactions on Signal
Processing}, vol.58, no.8, Aug. 2010, pp.4182-4194.

\bibitem{lei09}
L. Wang and R. C. de Lamare, "Low-Complexity Adaptive Step Size
Constrained Constant Modulus SG Algorithms for Blind Adaptive
Beamforming", \textit{Signal Processing}, vol. 89, no. 12, December
2009, pp. 2503-2513.

\bibitem{ccmavf}
L. Wang and R. C. de Lamare, ``Adaptive Constrained Constant Modulus
Algorithm Based on Auxiliary Vector Filtering for Beamforming," IEEE
Transactions on Signal Processing, vol. 58, no. 10, pp. 5408-5413,
Oct. 2010.


\bibitem{lei10}
L. Wang, R. C. de Lamare, M. Yukawa, "Adaptive Reduced-Rank
Constrained Constant Modulus Algorithms Based on Joint Iterative
Optimization of Filters for Beamforming," \textit{IEEE Transactions
on Signal Processing}, vol.58, no.6, June 2010, pp.2983-2997.

\bibitem{jio_ccm}
L. Wang, R. C. de Lamare and M. Yukawa, ``Adaptive reduced-rank
constrained constant modulus algorithms based on joint iterative
optimization of filters for beamforming", IEEE Transactions on
Signal Processing, vol.58, no. 6, pp. 2983-2997, June 2010.

\bibitem{ccmavf}
L. Wang and R. C. de Lamare, ``Adaptive constrained constant modulus
algorithm based on auxiliary vector filtering for beamforming", IEEE
Transactions on Signal Processing, vol. 58, no. 10, pp. 5408-5413,
October 2010.

\bibitem{stap_jio}
R. Fa and R. C. de Lamare, ``Reduced-Rank STAP Algorithms using
Joint Iterative Optimization of Filters," IEEE Transactions on
Aerospace and Electronic Systems, vol.47, no.3, pp.1668-1684, July
2011.

\bibitem{zhaocheng}
Z. Yang, R. C. de Lamare and X. Li, ``L1-Regularized STAP Algorithms
With a Generalized Sidelobe Canceler Architecture for Airborne
Radar," IEEE Transactions on Signal Processing, vol.60, no.2,
pp.674-686, Feb. 2012.

\bibitem{zhaocheng2}
Z. Yang, R. C. de Lamare and X. Li, ``Sparsity-aware space–time
adaptive processing algorithms with L1-norm regularisation for
airborne radar", IET signal processing, vol. 6, no. 5, pp. 413-423,
2012.

\bibitem{arh_eusipco}
Neto, F.G.A.; Nascimento, V.H.; Zakharov, Y.V.; de Lamare, R.C.,
"Adaptive re-weighting homotopy for sparse beamforming," in Signal
Processing Conference (EUSIPCO), 2014 Proceedings of the 22nd
European , vol., no., pp.1287-1291, 1-5 Sept. 2014

\bibitem{arh_taes}
Almeida Neto, F.G.; de Lamare, R.C.; Nascimento, V.H.; Zakharov,
Y.V.,``Adaptive reweighting homotopy algorithms applied to
beamforming," IEEE Transactions on Aerospace and Electronic Systems,
vol.51, no.3, pp.1902-1915, July 2015.

\bibitem{dfjio}
L. Wang, R. C. de Lamare and M. Haardt, ``Direction finding
algorithms based on joint iterative subspace optimization," IEEE
Transactions on Aerospace and Electronic Systems, vol.50, no.4,
pp.2541-2553, October 2014.

\bibitem{rdrab}
S. D. Somasundaram, N. H. Parsons, P. Li and R. C. de Lamare,
``Reduced-dimension robust capon beamforming using Krylov-subspace
techniques," IEEE Transactions on Aerospace and Electronic Systems,
vol.51, no.1, pp.270-289, January 2015.

\bibitem{dcg_conf}
S. Xu and R.C de Lamare, , \textit{Distributed conjugate gradient
strategies for distributed estimation over sensor networks}, Sensor
Signal Processing for Defense SSPD, September 2012.


\bibitem{dcg}
S. Xu, R. C. de Lamare, H. V. Poor, ``Distributed Estimation Over
Sensor Networks Based on Distributed Conjugate Gradient Strategies",
IET Signal Processing, 2016 (to appear).

\bibitem{dce}
S. Xu, R. C. de Lamare and H. V. Poor, \textit{Distributed
Compressed Estimation Based on Compressive Sensing}, IEEE Signal
Processing letters, vol. 22, no. 9, September 2014.

\bibitem{drr_conf}
S. Xu, R. C. de Lamare and H. V. Poor, ``Distributed reduced-rank
estimation based on joint iterative optimization in sensor
networks," in Proceedings of the 22nd European Signal Processing
Conference (EUSIPCO), pp.2360-2364, 1-5, Sept. 2014

\bibitem{dta_conf1}
S. Xu, R. C. de Lamare and H. V. Poor, ``Adaptive link selection
strategies for distributed estimation in diffusion wireless
networks," in Proc. IEEE International Conference onAcoustics,
Speech and Signal Processing (ICASSP),  , vol., no., pp.5402-5405,
26-31 May 2013.

\bibitem{dta_conf2}
S. Xu, R. C. de Lamare and H. V. Poor, ``Dynamic topology adaptation
for distributed estimation in smart grids," in Computational
Advances in Multi-Sensor Adaptive Processing (CAMSAP), 2013 IEEE 5th
International Workshop on , vol., no., pp.420-423, 15-18 Dec. 2013.

\bibitem{dta_ls}
S. Xu, R. C. de Lamare and H. V. Poor, ``Adaptive Link Selection
Algorithms for Distributed Estimation", EURASIP Journal on Advances
in Signal Processing, 2015.

\bibitem{song}
N. Song, R. C. de Lamare, M. Haardt, and M. Wolf, ``Adaptive Widely
Linear Reduced-Rank Interference Suppression based on the
Multi-Stage Wiener Filter," IEEE Transactions on Signal Processing,
vol. 60, no. 8, 2012.

\bibitem{wljio}
N. Song, W. U. Alokozai, R. C. de Lamare and M. Haardt, ``Adaptive
Widely Linear Reduced-Rank Beamforming Based on Joint Iterative
Optimization,"  IEEE Signal Processing Letters, vol.21, no.3, pp.
265-269, March 2014.

\bibitem{barc}
R.C. de Lamare, R. Sampaio-Neto and M. Haardt, "Blind Adaptive
Constrained Constant-Modulus Reduced-Rank Interference Suppression
Algorithms Based on Interpolation and Switched Decimation,"
\textit{IEEE Trans. on Signal Processing},  vol.59, no.2,
pp.681-695, Feb. 2011.

\bibitem{jiomber}
Y. Cai, R. C. de Lamare, ``Adaptive Linear Minimum BER Reduced-Rank
Interference Suppression Algorithms Based on Joint and Iterative
Optimization of Filters," IEEE Communications Letters, vol.17, no.4,
pp.633-636, April 2013.

\bibitem{saalt}
R. C. de Lamare and R. Sampaio-Neto, ``Sparsity-Aware Adaptive
Algorithms Based on Alternating Optimization and Shrinkage," IEEE
Signal Processing Letters, vol.21, no.2, pp.225,229, Feb. 2014.

\bibitem{alrdoa}
L. Qiu, Y. Cai, R. C. de Lamare and M. Zhao, ``Reduced-Rank DOA
Estimation Algorithms Based on Alternating Low-Rank Decomposition,"
in IEEE Signal Processing Letters, vol. 23, no. 5, pp. 565-569, May
2016.

\bibitem{mmimo}
R. C. de Lamare, ``Massive MIMO Systems: Signal Processing
Challenges and Future Trends", Radio Science Bulletin, December
2013.

\bibitem{wence}
W. Zhang, H. Ren, C. Pan, M. Chen, R. C. de Lamare, B. Du and J.
Dai, ``Large-Scale Antenna Systems With UL/DL Hardware Mismatch:
Achievable Rates Analysis and Calibration", IEEE Trans. Commun.,
vol.63, no.4, pp. 1216-1229, April 2015.

\bibitem{Costa}
M. Costa, "Writing on dirty paper," \textit{IEEE Trans. Inform.
Theory}, vol. 29, no. 3, pp. 439-441, May 1983.

\bibitem{delamare_ieeproc}
R. C. de Lamare and A. Alcaim, "Strategies to improve the
performance of very low bit rate speech coders and application to a
1.2 kb/s codec" IEE Proceedings- Vision, image and signal
processing, vol. 152, no. 1, February, 2005.

\bibitem{TDS_clarke}
P. Clarke and R. C. de Lamare, "Joint Transmit Diversity
Optimization and Relay Selection for Multi-Relay Cooperative MIMO
Systems Using Discrete Stochastic Algorithms," \emph{IEEE
Communications Letters}, vol.15, no.10, pp.1035-1037, October 2011.

\bibitem{TDS_2}
P. Clarke and R. C. de Lamare, "Transmit Diversity and Relay
Selection Algorithms for Multirelay Cooperative MIMO Systems"
\emph{IEEE Transactions on Vehicular Technology}, vol.61, no. 3, pp.
1084-1098, March 2012.

\bibitem{armo}
T. Peng, R. C. de Lamare and A. Schmeink, ``Adaptive Distributed
Space-Time Coding Based on Adjustable Code Matrices for Cooperative
MIMO Relaying Systems," \emph{IEEE Transactions on Communications},
vol. 61, no. 7, pp. 2692-2703, July 2013.

\bibitem{buffer}
T. Peng and R. C. de Lamare, ``Adaptive Buffer-Aided Distributed
Space-Time Coding for Cooperative Wireless Networks," \emph{ IEEE
Transactions on Communications}, vol. 64, no. 5, pp. 1888-1900, May
2016.

\bibitem{switch_int}
Y. Cai, R. C. de Lamare, and R. Fa, ``Switched Interleaving
Techniques with Limited Feedback for Interference Mitigation in
DS-CDMA Systems," IEEE Transactions on Communications, vol.59, no.7,
pp.1946-1956, July 2011.

\bibitem{switch_mc}
Y. Cai, R. C. de Lamare, D. Le Ruyet, ``Transmit Processing
Techniques Based on Switched Interleaving and Limited Feedback for
Interference Mitigation in Multiantenna MC-CDMA Systems," IEEE
Transactions on Vehicular Technology, vol.60, no.4, pp.1559-1570,
May 2011.

\bibitem{smce}
T. Wang, R. C. de Lamare, and P. D. Mitchell, ``Low-Complexity
Set-Membership Channel Estimation for Cooperative Wireless Sensor
Networks," IEEE Transactions on Vehicular Technology, vol.60, no.6,
pp.2594-2607, July 2011.

\bibitem{TongW}
T. Wang, R. C. de Lamare and A. Schmeink, "Joint linear receiver
design and power allocation using alternating optimization
algorithms for wireless sensor networks," \textit{IEEE Trans. on
Vehi. Tech.}, vol. 61, pp. 4129-4141, 2012.

\bibitem{jpais_iet}
R. C. de Lamare, ``Joint iterative power allocation and linear
interference suppression algorithms for cooperative DS-CDMA
networks", IET Communications, vol. 6, no. 13 , 2012, pp. 1930-1942.

\bibitem{TARMO}
T. Peng, R. C. de Lamare and A. Schmeink, ``Adaptive Distributed
Space-Time Coding Based on Adjustable Code Matrices for Cooperative
MIMO Relaying Systems'', \emph{IEEE Transactions on Communications},
vol. 61, no. 7, July 2013.

\bibitem{keke1}
K. Zu, R. C. de Lamare, ``Low-Complexity Lattice Reduction-Aided
Regularized Block Diagonalization for MU-MIMO Systems'', IEEE.
Communications Letters, Vol. 16, No. 6, June 2012, pp. 925-928.

\bibitem{kekecl}
K. Zu, R. C. de Lamare, ``Low-Complexity Lattice Reduction-Aided
Regularized Block Diagonalization for MU-MIMO Systems'', IEEE.
Communications Letters, Vol. 16, No. 6, June 2012.

\bibitem{keke2} K. Zu, R. C. de Lamare and M.
Haart, ``Generalized design of low-complexity block diagonalization
type precoding algorithms for multiuser MIMO systems", IEEE Trans.
Communications, 2013.

\bibitem{Tomlinson}
M. Tomlinson, "New automatic equaliser employing modulo arithmetic,"
\textit{Electronic Letters}, vol. 7, Mar. 1971.

\bibitem{dopeg_cl} C. T. Healy and R. C. de Lamare,
``Decoder-optimised progressive edge growth algorithms for the
design of LDPC codes with low error floors",  \textit{IEEE
Communications Letters}, vol. 16, no. 6, June 2012, pp. 889-892.

\bibitem{peg_bf_iswcs}
A. G. D. Uchoa, C. T. Healy, R. C. de Lamare, R. D. Souza, ``LDPC
codes based on progressive edge growth techniques for block fading
channels", \textit{Proc. 8th International Symposium on Wireless
Communication Systems (ISWCS)}, 2011, pp. 392-396.

\bibitem{gqcpeg}
A. G. D. Uchoa, C. T. Healy, R. C. de Lamare, R. D. Souza,
``Generalised Quasi-Cyclic LDPC codes based on progressive edge
growth techniques for block fading channels",  \textit{Proc.
International Symposium Wireless Communication Systems (ISWCS)},
2012, pp. 974-978.

\bibitem{peg_bf_cl}
A. G. D. Uchoa, C. T. Healy, R. C. de Lamare, R. D. Souza, ``Design
of LDPC Codes Based on Progressive Edge Growth Techniques for Block
Fading Channels", \textit{IEEE Communications Letters}, vol. 15, no.
11, November 2011, pp. 1221-1223.

\bibitem{memd_tcom}
C. T. Healy and R. C. de Lamare, ``Design of LDPC Codes Based on
Multipath EMD Strategies for Progressive Edge Growth", \emph{IEEE
Transactions on Communications}, 2016.

\bibitem{Harashima}
H. Harashima and H. Miyakawa, "Matched-transmission technique for
channels with intersymbol interference," \textit{IEEE Trans.
Commun.}, vol. 20, Aug. 1972.

\bibitem{mbthpc}
K. Zu, R. C. de Lamare and M. Haardt, ``Multi-branch
tomlinson-harashima precoding for single-user MIMO systems," in
Smart Antennas (WSA), 2012 International ITG Workshop on , vol.,
no., pp.36-40, 7-8 March 2012.

\bibitem{zuthp}
K. Zu, R. C. de Lamare and M. Haardt, ``Multi-Branch
Tomlinson-Harashima Precoding Design for MU-MIMO Systems: Theory and
Algorithms," IEEE Transactions on Communications, vol.62, no.3,
pp.939,951, March 2014.


\bibitem{rmbthp}
L. Zhang, Y. Cai, R. C. de Lamare and M. Zhao,  ``Robust Multibranch
Tomlinson–Harashima Precoding Design in Amplify-and-Forward MIMO
Relay Systems," IEEE Transactions on Communications, vol.62, no.10,
pp.3476,3490, Oct. 2014.

\bibitem{Hochwald}
B. Hochwald, C. Peel and A. Swindlehurst, "A vector-perturbation
technique for near capacity multiantenna multiuser communication -
Part II: Perturbation," \textit{IEEE Trans. Commun.}, vol. 53, no.
3, Mar. 2005.

\bibitem{BDVP}
C. B. Chae, S. Shim and R. W. Heath, "Block diagonalized vector
perturbation for multiuser MIMO systems," \textit{IEEE Trans.
Wireless Commun.}, vol. 7, no. 11, pp. 4051 - 4057, Nov. 2008.

\bibitem{delamare_mber}
R. C. de Lamare, R. Sampaio-Neto, ``Adaptive MBER decision feedback
multiuser receivers in frequency selective fading channels",
\textit{ IEEE Communications Letters}, vol. 7, no. 2, Feb. 2003, pp.
73 - 75.

\bibitem{rontogiannis}
A. Rontogiannis, V. Kekatos, and K. Berberidis," A Square-Root
Adaptive V-BLAST Algorithm for Fast Time-Varying MIMO Channels,"
\textit{IEEE Signal Processing Letters}, Vol. 13, No. 5, pp.
265-268, May 2006.

\bibitem{delamare_itic} R. C.
de Lamare, R. Sampaio-Neto, A. Hjorungnes, ``Joint iterative
interference cancellation and parameter estimation for CDMA
systems", \textit{IEEE Communications Letters}, vol. 11, no. 12,
December 2007, pp. 916 - 918.

\bibitem{stspadf}
Y. Cai and R. C. de Lamare, "Adaptive Space-Time Decision Feedback
Detectors with Multiple Feedback Cancellation", \textit{IEEE
Transactions on Vehicular Technology}, vol. 58, no. 8,  October
2009, pp. 4129 - 4140.

\bibitem{choi}
J. W. Choi, A. C. Singer, J Lee, N. I. Cho, ``Improved linear
soft-input soft-output detection via soft feedback successive
interference cancellation," \textit{IEEE Trans. Commun.}, vol.58,
no.3, pp.986-996, March 2010.


\bibitem{stbcccm}
R. C. de Lamare and R. Sampaio-Neto, ``Blind adaptive MIMO receivers
for space-time block-coded DS-CDMA systems in multipath channels
using the constant modulus criterion," IEEE Transactions on
Communications, vol.58, no.1, pp.21-27, January 2010.


\bibitem{FL11}
R. Fa, R. C. de Lamare, ``Multi-Branch Successive Interference
Cancellation for MIMO Spatial Multiplexing Systems", \textit{ IET
Communications}, vol. 5, no. 4, pp. 484 - 494, March 2011.

\bibitem{jio_mimo}
R.C. de Lamare and R. Sampaio-Neto, ``Adaptive reduced-rank
equalization algorithms based on alternating optimization design
techniques for MIMO systems," IEEE Trans. Veh. Technol., vol. 60,
no. 6, pp. 2482-2494, July 2011.

\bibitem{peng_twc} P. Li, R. C. de Lamare and R. Fa, ``Multiple
Feedback Successive Interference Cancellation Detection for
Multiuser MIMO Systems," \textit{IEEE Transactions on Wireless
Communications}, vol. 10, no. 8, pp. 2434 - 2439, August 2011.

\bibitem{spa}
R.C. de Lamare, R. Sampaio-Neto, ``Minimum mean-squared error
iterative successive parallel arbitrated decision feedback detectors
for DS-CDMA systems," IEEE Trans. Commun., vol. 56, no. 5, May 2008,
pp. 778-789.

\bibitem{spa2}
R.C. de Lamare, R. Sampaio-Neto, ``Minimum mean-squared error
iterative successive parallel arbitrated decision feedback detectors
for DS-CDMA systems," IEEE Trans. Commun., vol. 56, no. 5, May 2008.

\bibitem{jio_mimo} R.C. de Lamare and R. Sampaio-Neto, ``Adaptive
reduced-rank equalization algorithms based on alternating
optimization design techniques for MIMO systems," IEEE Trans. Veh.
Technol., vol. 60, no. 6, pp. 2482-2494, July 2011.

\bibitem{P.Li}
P. Li, R. C. de Lamare and J. Liu, ``Adaptive Decision Feedback
Detection with Parallel Interference Cancellation and Constellation
Constraints for Multiuser MIMO systems'', IET Communications, vol.7,
2012, pp. 538-547.

\bibitem{jingjing}
J. Liu, R. C. de Lamare, ``Low-Latency Reweighted Belief Propagation
Decoding for LDPC Codes," IEEE Communications Letters, vol. 16, no.
10, pp. 1660-1663, October 2012.

\bibitem{did}
P. Li and R. C. de Lamare, Distributed Iterative Detection With
Reduced Message Passing for Networked MIMO Cellular Systems, IEEE
Transactions on Vehicular Technology, vol.63, no.6, pp. 2947-2954,
July 2014.

\bibitem{bfidd}
A. G. D. Uchoa, C. T. Healy and R. C. de Lamare, ``Iterative
Detection and Decoding Algorithms For MIMO Systems in Block-Fading
Channels Using LDPC Codes," IEEE Transactions on Vehicular
Technology, 2015.

\bibitem{mbdf}
R. C. de Lamare, ``Adaptive and Iterative Multi-Branch MMSE Decision
Feedback Detection Algorithms for Multi-Antenna Systems", \emph{IEEE
Trans. Wireless Commun.}, vol. 14, no. 10, October 2013.


}
}
\end{thebibliography}
\end{document}